\documentclass[a4paper,11pt]{article}
\pdfoutput=1 

\usepackage{jcappub} 

\usepackage[utf8]{inputenc}
\usepackage[T1]{fontenc}
\usepackage{lmodern}
\usepackage{amsmath, amssymb, amsfonts} 
\usepackage{mathtools}
\usepackage{graphicx}       
	\graphicspath{ {./figures/} }
\usepackage[caption=false]{subfig} 
\usepackage[dvipsnames]{xcolor}   
\usepackage{physics}
\usepackage{tensor}
\usepackage{mathrsfs}
\usepackage{bm}
\usepackage{enumitem}
\usepackage{lipsum}

\def \eqdef {\coloneqq}
\def \be {\begin{equation}}
\def \ee {\end{equation}}
\def \nn {\nonumber}
\def \del {\partial}

\def \detOm {\abs{\bm{\Omega}}}  
\def \tildx {\tilde{X}}
\def \tildL {\tilde{\mathcal{L}}_m}  
\def \zetaeq {\zeta_{\mathrm{eq}}}  
\def \ycal {\mathcal{Y}}
\def \Sigmacr {\Sigma_{\mathrm{cr}}}  
\def \rph {r_{\mathrm{ph}}} 

\numberwithin{equation}{section}

\begin{document}
\title{Generating Rotating Spacetime in Ricci-Based Gravity: Naked Singularity as a Black Hole Mimicker}

\author[a,b]{Wei-Hsiang Shao,}
\author[c]{Che-Yu Chen}
\author[a,b,d]{and Pisin Chen}

\affiliation[a]{Department of Physics and Center for Theoretical Physics, \\National Taiwan University, Taipei 10617, Taiwan}
\affiliation[b]{Leung Center for Cosmology and Particle Astrophysics, \\National Taiwan University, Taipei 10617, Taiwan}
\affiliation[c]{Institute of Physics, Academia Sinica, \\Taipei 11529, Taiwan}
\affiliation[d]{Kavli Institute for Particle Astrophysics and Cosmology, SLAC National Accelerator Laboratory, Stanford University, Stanford, CA 94305, USA}

\emailAdd{whsshao@gmail.com}
\emailAdd{b97202056@gmail.com}
\emailAdd{pisinchen@phys.ntu.edu.tw}

\abstract{ 
Motivated by the lack of rotating solutions sourced by matter in General Relativity as well as in modified gravity theories, we extend a recently discovered exact rotating solution of the minimal Einstein-scalar theory to its counterpart in Eddington-inspired Born-Infeld gravity coupled to a Born-Infeld scalar field. 
This is accomplished with the implementation of a well-developed mapping between solutions of Ricci-Based Palatini theories of gravity and General Relativity. 
The new solution is parametrized by the scalar charge and the Born-Infeld coupling constant apart from the mass and spin of the compact object. 
Compared to the spacetime prior to the mapping, we find that the high-energy modifications at the Born-Infeld scale are able to suppress but not remove the curvature divergence of the original naked null singularity. 
Depending on the sign of the Born-Infeld coupling constant, these modifications may even give rise to an additional timelike singularity exterior to the null one. 
In spite of that, both of the naked singularities before and after the mapping are capable of casting shadows, and as a consequence of the mapping relation, their shadows turn out to be identical as seen by a distant observer on the equatorial plane.
Even though the scalar field induces a peculiar oblateness to the appearance of the shadow with its left and right endpoints held fixed, the closedness condition for the shadow contour sets a small upper bound on the absolute value of the scalar charge, which leads to observational features of the shadow closely resembling those of a Kerr black hole.
}


\arxivnumber{2011.07763}

\maketitle

\section{Introduction}
The conservation of angular momentum plays an extremely important role in many aspects of physics. 
In particular, at astrophysical scales, a direct consequence of angular momentum conservation is that almost all stellar objects that have been formed via gravitational collapse, such as stars and black holes, carry nonzero spins. 
Typically, an accurate physical description of those stellar objects that are endowed with a strong gravitational field should be based on Einstein's General Relativity (GR). 
However, due to the substantial complexity and nonlinearity of the Einstein equations, finding the solutions describing such rotating spacetime configurations is a daunting challenge. 
A few tricks have been put forward in an attempt to include angular momentum into a general static solution, the most renowned one being the Newman-Janis algorithm~\cite{1965JMP.....6..915N} (see~\cite{Drake:1998gf} or~\cite{Erbin:2016lzq} for a comprehensive description) that was successful in ``rederiving'' the Kerr solution using the Schwarzschild metric as its seed. 
Yet the algorithm hinges entirely on \emph{off-shell} manipulations, thus the field equations are not guaranteed to be preserved, and this operation does not necessarily reproduce the expected results~\cite{Pirogov:2013wia, Hansen:2013owa}.

A more precise disclosure of the nature of these compact objects urges us to not just settle for vacuum solutions but to also take into account the couplings between gravity and matter degrees of freedom.
Indeed, besides the well-studied electrically charged black holes in the Einstein-Maxwell theory, the inclusion of scalar fields has been shown to give rise to boson stars~\cite{Schunck:2003kk, Macedo:2013jja, Liebling:2012fv}, hairy black holes~\cite{Herdeiro:2015waa}, dilaton effects in black holes~\cite{Boulware:1986dr, Gibbons:1987ps, Garfinkle:1990qj, Holzhey:1991bx, Gregory:1992kr, Rakhmanov:1993yd}, spontaneous scalarization~\cite{Damour:1993hw, Damour:1996ke},\footnote{
The scalar-tensor setup of gravity has sparked interest due to the natural emergence of scalar fields in supergravity~\cite{VanNieuwenhuizen:1981ae} and in the low-energy limit of string theory~\cite{green1987superstring}.  
}
and even structural changes in the Cauchy horizon as well as the central singularity of both Kerr and Reissner-Nordstr\"om black holes~\cite{Chesler:2019pss, Chesler:2019tco}.
The simplest among these models is GR with a minimally coupled real massless scalar field (the minimal Einstein-scalar theory, or MES theory for short) which provides additional long-range effects. 
Several static solutions of the MES theory have been found in the past, including the spherically symmetric FJNW solution obtained by Fisher, Janis, Newman, and Winicour~\cite{Fisher:1948yn, Janis:1968zz} (and later by Wyman~\cite{Wyman:1981bd}),\footnote{
The FJNW solution and the Wyman solution were shown to be the same~\cite{Virbhadra:1997ie}. 
}
and the scalar-deformed Zipoy-Voorhees (ZV)~\cite{1966JMP.....7.1137Z, Voorhees:1971wh} and Erez-Rosen~\cite{erez1959} metrics presented in~\cite{Turimov:2018guy}. 
These solutions have in common a dimensionless scalar charge parameter which characterizes the effective coupling strength of the gravitating scalar field. 
Rotating solutions of the theory have also been discovered, and one of them is a Kerr-like rotating generalization of the scalar-deformed ZV metric~\cite{Bogush:2020lkp} that will be taken as the starting point for this work. 
This rotating MES solution carries oblate deformation due to the nonvanishing multipole moments induced by the scalar field, so even in the static limit it departs from spherical symmetry in a way similar to the vacuum ZV metric~\cite{1966JMP.....7.1137Z, Voorhees:1971wh}. 
An interesting feature often shared by asymptotically flat spacetimes with a massless scalar source is the presence of a naked singularity~\cite{Chase:1970}, and the rotating MES solution is no exception. 
In fact, it possesses a curvature singularity at the surface where the event horizon of the Kerr metric is located. 

The point of emphasis is that without considering any simplifying ansatz, perturbative treatments of slowly rotating scenarios, or numerical approaches, constructing the rotating counterpart of a non-vacuum solution remains a nontrivial task in GR. 
Take for instance the FJNW solution, naive application of the Newman-Janis algorithm~\cite{PhysRevD.31.1280} leads to a line element that fails to fulfill the field equations~\cite{Pirogov:2013wia, Bogush:2020lkp}. 
More sophisticated methods have to be conjured up in order to arrive at exact rotating solutions (see, e.g.,~\cite{Chauvineau:2018zjy, Bogush:2020lkp} and references therein), such as the Cl\'ement transformation~\cite{Clement:1997tx} that assisted in the discovery of the aforementioned rotating MES solution.

To make matters worse, numerous unsolved mysteries in the Universe, such as the dark energy and dark matter puzzles, the inevitable occurrence of spacetime singularities, and the inability to consistently incorporate GR and quantum theory, seem to thrust upon us the possibility of alternative theories of gravity~\cite{Capozziello:2011et, Nojiri:2017ncd}.
Needless to say, these modified gravity theories contain more complicated structures, and their vacuum solutions are even more beyond reach than the ones in GR, let alone the non-vacuum rotating solutions.  
On the other hand, in the wake of the recent imaging of the M87 galactic center by the Event Horizon Telescope (EHT) Collaboration~\cite{Akiyama:2019cqa} and the detection of gravitational waves by the LIGO/Virgo Collaboration (see for example~\cite{Abbott:2016blz, Abbott:2017oio}), a promising new era has just begun in which a large number of modified gravity theories along with their spacetime solutions can be put to test as observations probe deeper into the strong gravity regime~\cite{Berti:2015itd, Yagi:2016jml}.  
That being said, without precise knowledge of the gravitational field configurations produced in these theories, there will still be difficulties in surveying the details of their predictions to the full extent.
Therefore, efficient methods designed to obtain exact analytic (rotating) solutions would be highly desirable from a theoretical standpoint.

In recent years, a particular class of gravitational theories termed Ricci-Based Gravity (RBG) theories~\cite{Afonso:2018bpv} has caught considerable attention, as it has stimulated thoughts on pursuing exact solutions of RBGs and also GR in a systematic way. 
This class of theories is formulated assuming that the metric and the connection are independent (Palatini formulation), and the gravitational Lagrangian is constructed out of contractions between the metric and the \emph{symmetrized} Ricci tensor, with the latter defined in terms of the affine connection.
Besides GR itself, the family of RBGs encompasses quite a variety of modified gravity theories that have made their appearances in the literature, including Palatini $f(R)$ gravity~\cite{Olmo:2011uz}, quadratic Palatini gravity~\cite{Borowiec:1996kg}, Eddington-inspired Born-Infeld (EiBI) gravity~\cite{Banados:2010ix}, and other extensions of Born-Infeld gravity theories~\cite{Makarenko:2014lxa, Odintsov:2014yaa, Jimenez:2014fla, Chen:2015eha}. 
These theories modify GR at high energy scales, and their impact on cosmological and astrophysical scenarios have been widely investigated~\cite{Allemandi:2004wn, Li:2007xw, Olmo:2011np, Olmo:2012nx, Lobo:2013adx, Lobo:2014zla, BeltranJimenez:2017doy}.

Given that only the symmetric part of the Ricci tensor contributes to the action, RBGs thus possess projective symmetry.
They have the merit of avoiding ghost instabilities that higher-order gravity theories are generally prone to, since projective symmetry ensures no propagating degrees of freedom accompany the connection~\cite{Afonso:2017bxr} (see also~\cite{BeltranJimenez:2019acz, Jimenez:2020dpn}). 
The connection field is non-dynamical, so RBGs in vacuum are in essence nothing but the Palatini formulation of GR, which allows these theories to naturally lie within the constraints of the speed of gravitational waves inferred from observations~\cite{Monitor:2017mdv, Cornish:2017jml}. 
When bosonic matter fields involving possible nonminimal couplings to gravity are included, the connection can be integrated out to arrive at the Einstein frame representation of RBGs where the matter sector contains new interactions and is now coupled to an auxiliary Einstein-frame metric~\cite{Afonso:2017bxr}.
The auxiliary metric is related to the physical spacetime metric in the RBG frame via an algebraic mapping, and its metric field equations are formally equivalent to the Einstein equations sourced by the stress-energy tensor of the matter sector in the Einstein frame. 
From a purely mathematical point of view, the mapping relation then connects a solution of RBG coupled to some matter source with that of GR coupled to the same matter source but described by different interactions. 
Hence, rather than tackling the field equations directly, one can exploit this mapping to find classical solutions on one side simply by borrowing from the pool of solutions on the other side. 
Such a correspondence between RBGs and GR has been well established for cases where the matter sector consists of scalar fields~\cite{Afonso:2018hyj}, fluids~\cite{Afonso:2018bpv}, and electromagnetic fields~\cite{Delhom:2019zrb}. 
Furthermore, the mapping has been shown to be successful in generating solutions of different RBGs~\cite{Afonso:2019fzv, Olmo:2020fnk}, with a more recent and intriguing example being a rotating charged black hole in EiBI gravity coupled to Born-Infeld electrodynamics~\cite{Guerrero:2020azx} obtained from its counterpart, the Kerr-Newman solution, in GR. 

This work is based on the mapping between RBGs and GR with a scalar source, and the goal is to extend the newly discovered rotating naked singularity solution of MES to its counterpart solution within the framework of an RBG theory. 
More specifically, the RBG theory that we will be aiming for with the mapping is EiBI gravity, which is motivated due to its ability to ameliorate spacetime singularities in the early Universe as well as a handful of those contained in astrophysical objects~\cite{Banados:2010ix, Pani:2011mg, Pani:2012qb, Avelino:2012ue, Cho:2012vg, Scargill:2012kg, Olmo:2013gqa}. 
We will start with the rotating MES solution as the seed and implement the mapping to obtain a new exact rotating solution of EiBI gravity.
As we shall see, this mapping brings the free canonical massless scalar field on the GR side to a scalar field modified by the nonlinear Born-Infeld corrections in the matter sector on the EiBI side. 
Similar to the structure of EiBI gravity, the modifications to the matter sector can be thought of as effectively an infinite series of higher derivative interactions characterized by the EiBI mass scale. 
Their ramifications are represented by the EiBI coupling constant which enters into the generated spacetime solution as an additional parameter aside from the mass, the spin, and the scalar charge. 
With regard to the generated solution, it will be shown that the naked singularity in the original MES solution is carried over to the EiBI side and lies at the \emph{would-be} event horizon of the new spacetime (a null singularity).
The corrections introduced in accordance with the Born-Infeld prescription can at most tame the curvature divergence of the naked singularity to become milder but are incapable of fully resolving it. 
Not only that, if the EiBI coupling constant is positive, these high-energy corrections trigger an additional divergent behavior of the stress-energy tensor of the scalar matter on a timelike hypersurface outside the previous null singularity.
The abovementioned singularities in the solution are \emph{true} singularities in the sense that they result in geodesic incompleteness of the spacetime manifold~\cite{Geroch:1968ut, Hawking:1969sw}.\footnote{
See~\cite{Olmo:2015bya} for an explicit example in which spacetimes plagued with curvature divergences are geodesically complete. 
}
That is, causal geodesics can reach these singularities in finite affine parameters but are obstructed from extending further beyond them. 
Although the formation of naked singularities are hypothesized to be forbidden by the cosmic censorship conjecture~\cite{Penrose:1969pc}, numerous studies have provided counterexamples demonstrating that it is possible for such singularities to take place as the end product of gravitational collapse under suitable initial conditions~\cite{Shapiro:1991zza, Joshi:1993zg, Harada:1998cq, Joshi:2001xi, Joshi:2011zm, Joshi:2011qq, Banerjee:2017njk, Bhattacharya:2017chr, Mosani:2020ena}.

As is widely known, hosting a naked singularity does not prevent the object from showcasing interesting optical properties~\cite{Virbhadra:2002ju, Nakao:2002kc, Virbhadra:2007kw, Gyulchev:2008ff, Bambi:2008jg, Kovacs:2010xm, Sahu:2012er, Joshi:2013dva, Kong:2013daa, Ortiz:2014kpa, Ortiz:2015rma, Shaikh:2018lcc, Gyulchev:2019tvk, Shaikh:2019hbm, Dey:2020haf, Joshi:2020tlq, Dey:2020bgo}, some of which nonetheless agree with observations to a high degree. 
In fact, we shall see that both the rotating MES spacetime and the new EiBI spacetime have their naked singularities cloaked by photon regions which act as potential barriers for impacting photons.
The photon region defines a boundary in the sky of the observer that encloses an area which is essentially the cross section for capturing photons~\cite{Cunha:2018acu}, and the existence of it enables the two naked singularities to cast shadows.
Their apparent shadow images will be investigated, and under certain situations they are identical, namely that the appearance is insensitive to the Born-Infeld modifications. 
This can be traced back to the mapping relation that ties together the two solutions, as we will explore later.
The peculiar oblateness of the shadow induced by the scalar field, as well as a spin-dependent upper bound on the scalar charge parameter imposed by observational consistency, will also be discussed.
Most importantly, we will highlight the astrophysical significance of both solutions by showing that their shadow contours bear close resemblance to that of a Kerr black hole with the same amount of mass and spin, thus displaying early signs of them being possible candidates for black hole mimickers. 

The paper is organized as follows. 
We begin our discussion in section~\ref{sec:2} by recapping the main elements that lay the foundations of the recipe for generating solutions, including the Einstein frame representation of the field equations of RBGs, the form of the mapping with the matter source being a scalar, followed by an explicit application of it to EiBI gravity.
In section~\ref{sec:new soln}, we briefly depict the rotating solution of the MES theory that was discovered recently, and go on to construct its counterpart in the EiBI framework through the mapping.
In section~\ref{sec:properties}, we examine the spacetime properties of the generated EiBI solution, with a subsection devoted to a primitive study of the shadow cast by it.
The summary and several remarks on prospective directions for future research follow in section~\ref{sec:conclusion}.
The convention $c = \kappa^2 = 1$ for the units will be used throughout this paper, where $c$ is the speed of light, and $\kappa^2 \equiv 8\pi G / c^4$ is the Einstein gravitational constant. 
We will also adopt the mostly plus signature $(-, +, +, +)$ for the metrics. 

\section{Theoretical framework}
\label{sec:2}

\subsection{Ricci-Based Gravity theories}
\label{sec:RBG intro}
Ricci-Based Gravity (RBG)~\cite{Afonso:2018bpv} is a particular family of gravitational theories in which the full action of the form
\be 
\label{eq:RBG action}
\mathcal{S} = \int d^4x \, \sqrt{-g} \, \mathcal{L}_G \big[ g^{\mu\nu}, \mathcal{R}_{(\mu\nu)}(\Gamma) \big] + \mathcal{S}_m[g_{\mu\nu}, \Psi]
\ee
is considered in the Palatini formulation (with the metric $g_{\mu\nu}$ and the affine connection $\Gamma_{\mu\nu}^{\lambda}$ treated as independent field variables).
The gravitational Lagrangian $\mathcal{L}_G [ g^{\mu\nu}, \mathcal{R}_{(\mu\nu)}(\Gamma) ]$ is a scalar function built out of contractions involving the inverse spacetime metric $g^{\mu\nu}$ and the \emph{symmetrized} Ricci tensor $\mathcal{R}_{(\mu\nu)}(\Gamma)$, which is constructed solely from the independent connection $\Gamma_{\mu\nu}^{\lambda}$, i.e. $\mathcal{R}_{\mu\nu} = \mathcal{R}\indices{^{\alpha}_{\mu\alpha\nu}}$, where the Riemann tensor is defined as
\be 
\mathcal{R}\indices{^{\rho}_{\sigma\mu\nu}}(\Gamma) = \del_{\mu} \Gamma_{\nu\sigma}^{\rho} - \del_{\nu} \Gamma_{\mu\sigma}^{\rho} + \Gamma_{\mu\alpha}^{\rho} \Gamma_{\nu\sigma}^{\alpha} - \Gamma_{\nu\alpha}^{\rho} \Gamma_{\mu\sigma}^{\alpha} \, .
\ee
As for the matter sector, we consider a matter action $\mathcal{S}_m[g_{\mu\nu}, \Psi]$ which depends only on the metric and minimally coupled matter fields collectively represented by $\Psi$. 

By performing independent variations of the action \eqref{eq:RBG action} with respect to the metric and the connection respectively, we obtain the following two field equations \cite{Afonso:2017bxr}:
\begin{align}
\label{eq:RBG eom1}
2 \, \pdv{\mathcal{L}_G}{g^{\mu\nu}} - \mathcal{L}_G \, g_{\mu\nu} &= T_{\mu\nu} \, , \\
\label{eq:RBG eom2}
\nabla^{(\Gamma)}_{\lambda} \big[ \sqrt{-q} \, q^{\mu\nu} \big] - \delta_{\lambda}^{\mu} \, \nabla^{(\Gamma)}_{\alpha} \big[ \sqrt{-q} \, q^{\alpha\nu} \big] &= \sqrt{-q} \, \big[ \mathcal{T}\indices{^{\mu}_{\lambda\alpha}} q^{\alpha\nu} + \mathcal{T}\indices{^{\alpha}_{\alpha\lambda}} q^{\mu\nu} - \delta_{\lambda}^{\mu} \, \mathcal{T}\indices{^{\alpha}_{\alpha\beta}} q^{\beta\nu} \big] \, ,
\end{align}
where $T_{\mu\nu} = \frac{-2}{\sqrt{-g}} \frac{\delta \mathcal{S}_m}{\delta g^{\mu\nu}}$ is the stress-energy tensor of matter, $\nabla_{\mu}^{(\Gamma)}$ is the covariant derivative associated with $\Gamma_{\mu\nu}^{\lambda}$, and $\mathcal{T}\indices{^{\lambda}_{\mu\nu}} = 2\Gamma_{[\mu\nu]}^{\lambda}$ is the torsion tensor.
In the above equations, we have also introduced the \emph{auxiliary} metric $q_{\mu\nu}$ defined by~\cite{Afonso:2018hyj}
\be
\label{eq:q defn}
\sqrt{-q} \, q^{\mu\nu} \equiv 2 \sqrt{-g} \, \pdv{\mathcal{L}_G}{\mathcal{R}_{(\mu\nu)}} \, ,
\ee
with $q$ being its determinant. 
There is no hypermomentum sourcing the connection field equations \eqref{eq:RBG eom2} due to the matter fields not coupling to the connection.\footnote{
We will set aside the case of spinor fields and deal with just minimally coupled bosonic fields in this work.
}
Moreover, since the action \eqref{eq:RBG action} contains only the symmetric part of the Ricci tensor, the RBGs possess projective invariance.
This allows us to be able to gauge away the torsion by a suitable choice of projective transformation~\cite{Afonso:2017bxr}, whereas the (symmetric part of the) connection is given by the Levi-Civita connection of the auxiliary metric $q_{\mu\nu}$. 
In other words, the connection which solves eq.~\eqref{eq:RBG eom2} can be written as
\be 
\label{eq:RBG connec}
\Gamma_{\mu\nu}^{\lambda} = \frac{1}{2} \, q^{\lambda\alpha} (\del_{\mu} q_{\nu\alpha} + \del_{\nu} q_{\mu\alpha} - \del_{\alpha} q_{\mu\nu}) \, ,
\ee
with $\mathcal{T}\indices{^{\lambda}_{\mu\nu}} = 0$. 
Though $q_{\mu\nu}$ depends generally on the Ricci tensor $\mathcal{R}_{(\mu\nu)}$, which itself is defined in terms of the connection, we can remove this dependence by resorting to the metric field equations \eqref{eq:RBG eom1} and express $q_{\mu\nu}$ in terms of the spacetime metric $g_{\mu\nu}$ and the stress-energy tensor $T_{\mu\nu}$. 
The affine connection~\eqref{eq:RBG connec} is then completely determined by algebraic equations and thus carries no dynamical degrees of freedom. 
This crucial feature stemming from projective invariance ensures the absence of ghosts in these theories~\citep{BeltranJimenez:2019acz}.
Before moving forward, it is worth pointing out that the auxiliary metric $q_{\mu\nu}$ is introduced solely for mathematical convenience. 
Dynamics of the RBG theory~\eqref{eq:RBG action} are still governed by the spacetime metric $g_{\mu\nu}$ along with its Levi-Civita connection (rather than the independent connection $\Gamma_{\mu\nu}^{\lambda}$) that defines the true covariant derivative of the geometry~\cite{Sotiriou:2008rp}, and the actual physical quantities are the ones that are associated with them, such as $R(g) \eqdef g^{\mu\nu} R_{\mu\nu}(g)$ and $R_{\mu\nu}(g)R^{\mu\nu}(g)$.\footnote{
Notice the distinction between the symbols used for curvature quantities of the physical metric: $R_{\mu\nu}(g)$, $R(g)$, etc., and those of the auxiliary metric: $\mathcal{R}_{\mu\nu}(q)$, $\mathcal{R}(q) \eqdef q^{\mu\nu} \mathcal{R}_{\mu\nu}(q)$, etc.
}

It is useful to introduce the deformation matrix $\Omega\indices{^{\mu}_{\nu}}$ which links $g_{\mu\nu}$ and the auxiliary metric $q_{\mu\nu}$ through the relation
\be 
\label{eq:Om defn}
q_{\mu\nu} = g_{\mu\alpha} \, \Omega\indices{^{\alpha}_{\nu}} \, .
\ee
Given a specific RBG theory, the deformation matrix can be worked out from eq.~\eqref{eq:q defn}. 
Now, the fact that the connection is Levi-Civita with respect to the auxiliary metric $q_{\mu\nu}$ guarantees the existence of an Einstein frame representation for the action \eqref{eq:RBG action}, with new interactions in the matter sector coupled to $q_{\mu\nu}$~\cite{Afonso:2017bxr}. 
At the level of the field equations, one finds that eq.~\eqref{eq:RBG eom1} can be brought to the form~\cite{Delhom:2019zrb}
\be 
\label{eq:Ein_like}
\mathcal{G}\indices{^{\mu}_{\nu}}(q) = \frac{1}{\sqrt{\detOm}} \Bigg[ T\indices{^{\mu}_{\nu}} - \left( \mathcal{L}_G + \frac{T}{2} \right) \delta\indices{^{\mu}_{\nu}} \Bigg] \, ,
\ee
where $\mathcal{G}\indices{^{\mu}_{\nu}}(q) \equiv q^{\mu\alpha} \mathcal{G}_{\alpha\nu}(q) = q^{\mu\alpha} \big( \mathcal{R}_{\alpha\nu}(q) - \frac{1}{2} \, q_{\alpha\nu} \mathcal{R}(q) \big)$ is the Einstein tensor of the auxiliary metric $q_{\mu\nu}$, $\detOm$ denotes the determinant of the matrix $\Omega\indices{^{\mu}_{\nu}}$, and $T = g^{\mu\nu} T_{\mu\nu}$ is the trace of the stress-energy tensor.
As mentioned, with $\mathcal{L}_G$ and $\Omega\indices{^{\mu}_{\nu}}$ being functions of contractions between $g_{\mu\nu}$ and $\mathcal{R}_{(\mu\nu)}$ in general, they can be algebraically related to the matter content, and the right-hand side of eq.~\eqref{eq:Ein_like} can be written \emph{on-shell} purely in terms of $T\indices{^{\mu}_{\nu}}$.
Therefore, eq.~\eqref{eq:Ein_like} is equivalent to the metric field equations (for the auxiliary metric) of the Palatini formulation of GR with a modified matter sector whose stress-energy tensor is given by the right-hand side of the equation. 
Written in this form, it is now also obvious that eq.~\eqref{eq:Ein_like} in vacuum is consistent with the earlier statement that these RBG theories propagate no additional degrees of freedom in the gravitational sector other than the two massless tensorial ones. 

\subsection{Mapping between RBGs and GR with a scalar}
\label{sec:map}
The Einstein frame representation \eqref{eq:Ein_like} of the metric field equations of RBGs with minimally coupled matter fields suggests the idea of identifying it as the Einstein field equations of GR, provided that we reinterpret the right-hand side of it as a modified stress-energy tensor $\tilde{T}\indices{^{\mu}_{\nu}} \eqdef q^{\mu\alpha} \tilde{T}_{\alpha\nu}$ that is minimally coupled to the auxiliary metric $q_{\mu\nu}$.\footnote{
The tilde symbol will be used throughout this paper to denote quantities in the Einstein frame associated with the auxiliary metric.
}
Note that at this point the right-hand side of eq.~\eqref{eq:Ein_like} still contains the spacetime metric $g_{\mu\nu}$, which generally appears in $T\indices{^{\mu}_{\nu}}$. 
However, by the field redefinition \eqref{eq:Om defn}, we can always replace $g_{\mu\nu}$ in favor of $q_{\mu\nu}$ and the matter fields such that the right-hand side of eq.~\eqref{eq:Ein_like} takes on the same structure as $\tilde{T}\indices{^{\mu}_{\nu}}$, i.e. 
\be 
\label{eq:mapping eq}
\tilde{T}\indices{^{\mu}_{\nu}}(q_{\mu\nu}, \Psi) = \frac{1}{\sqrt{\detOm}} \Bigg[ T\indices{^{\mu}_{\nu}} - \bigg( \mathcal{L}_G + \frac{T}{2} \bigg) \delta\indices{^{\mu}_{\nu}} \Bigg] \, .
\ee 
This algebraic equation relates the matter sectors in the Einstein frame and the original RBG frame.
Once the correspondence between the two frames is established, the field equations $\mathcal{G}\indices{^{\mu}_{\nu}}(q) = \tilde{T}\indices{^{\mu}_{\nu}}$ of the RBG theory will become mathematically equivalent to a problem in the framework of GR, and eq.~\eqref{eq:Om defn} would then serve as a portal for us to correlate spacetime solutions of RBGs and GR or, better yet, obtain solutions of RBGs from known solutions of GR (and vice versa). 
This idea has been implemented in previous works for fluids~\cite{Afonso:2018bpv}, scalar fields~\cite{Afonso:2018hyj, Afonso:2019fzv}, and electromagnetic fields~\cite{Delhom:2019zrb, Guerrero:2020azx, Olmo:2020fnk}.

Let us briefly review the correspondence between the matter sectors of the two frames presented in~\cite{Afonso:2018hyj} for the case of a single scalar field. We refer the reader to appendix~\ref{app:map} for a detailed derivation.
The starting point is the RBG theory~\eqref{eq:RBG action} with the matter action given by
\be
\label{eq:scalar ac}
\mathcal{S}_m(X, \phi) = -\frac{1}{2} \int d^4x \, \sqrt{-g} \, \mathcal{L}_m(X, \phi) \, ,
\ee
which describes a general minimally coupled non-canonical scalar field. 
The Lagrangian density $\mathcal{L}_m$ is some arbitrary function of its arguments, where $X$ is the trace of
\be
\label{eq:X defn}
X\indices{^{\mu}_{\nu}} \equiv g^{\mu\alpha} \del_{\alpha} \phi \, \del_{\nu} \phi \, .
\ee
The stress-energy tensor of this scalar matter source reads
\be
\label{eq:T}
T\indices{^{\mu}_{\nu}} = \big( \del_X \mathcal{L}_m \big) X\indices{^{\mu}_{\nu}} - \frac{1}{2} \, \mathcal{L}_m(X, \phi) \delta\indices{^{\mu}_{\nu}} \, .
\ee
The other side of the correspondence is GR with a minimally coupled scalar field defined by the matter action of the RBG theory in the Einstein frame representation, which we write as
\be
\label{eq:tild scalar ac}
\tilde{\mathcal{S}}_m(\tilde{X}, \phi) = -\frac{1}{2} \int d^4x \, \sqrt{-q} \, \tilde{\mathcal{L}}_m(\tilde{X}, \phi) \, .
\ee
The modified matter Lagrangian $\tildL(\tildx, \phi)$ is coupled to the auxiliary metric $q_{\mu\nu}$, where, analogous to \eqref{eq:X defn} in the RBG frame, we have defined $\tilde{X}\indices{^{\mu}_{\nu}} \equiv q^{\mu\alpha} \partial_{\alpha} \phi \, \partial_{\nu} \phi$ in the Einstein frame.
As in \eqref{eq:T}, the stress-energy tensor of the scalar in the Einstein frame has the form $\tilde{T}\indices{^{\mu}_{\nu}} = (\del_{\tilde{X}} \tilde{\mathcal{L}}_m)  \tilde{X}\indices{^{\mu}_{\nu}} - (\tildL \delta\indices{^{\mu}_{\nu}})/2$.
Now, in order for the correspondence to be self-consistent, not only does the algebraic equation \eqref{eq:mapping eq} have to be satisfied, one also has to make sure that the scalar field solution is compatible with the field equations of both $\mathcal{S}_m$ and $\mathcal{\tilde{S}}_m$. 
As discussed in appendix~\ref{app:map}, this leads to the following necessary relations between the matter Lagrangian densities in the two frames: 
\begin{align} 
\label{eq:map_L}
\tildL(\tildx, \phi) &= \frac{1}{\sqrt{\detOm}} \, \big( 2 \mathcal{L}_G + X \del_X \mathcal{L}_m - \mathcal{L}_m \big) \, , \\
\label{eq:map_x tildx}
\tildx \, \del_{\tildx} \tildL &= \frac{1}{\sqrt{\detOm}} \, X \, \del_X \mathcal{L}_m \, , \\
\label{eq:map_dphi}
\del_{\phi} \tildL &= \frac{1}{\sqrt{\detOm}} \, \del_{\phi} \mathcal{L}_m \, ,
\end{align}
which are key to the mapping, as they allow us to construct the matter Lagrangian density $\tildL$ in the Einstein frame in terms of quantities in the RBG frame. 

\subsection{Eddington-inspired Born-Infeld gravity}
\label{sec:EiBI map}
Let us now dive into a specific case of the mapping where the RBG under consideration is the Eddington-inspired Born-Infeld (EiBI) gravity theory~\cite{Banados:2010ix} (see~\cite{BeltranJimenez:2017doy} for a thorough review). 
The action for the EiBI gravity theory is given by
\be 
\label{eq:eibi ac}
\mathcal{S}_{\text{EiBI}} = \frac{1}{\epsilon} \int d^4x \left[ \sqrt{-\abs*{g_{\mu\nu} + \epsilon \mathcal{R}_{(\mu\nu)}}} - \lambda \sqrt{-g} \right] \, ,
\ee
where $\epsilon$ is a constant length squared parameter associated with the additional Born-Infeld mass scale at which large-curvature corrections become relevant. 
At small curvature scales $\abs{\mathcal{R}_{\mu\nu}} \ll 1 / \epsilon$, \eqref{eq:eibi ac} reduces to the Einstein-Hilbert action with an effective cosmological constant given by $\Lambda = (\lambda-1)/\epsilon$.\footnote{
Although we can set $\lambda = 1$ since we are aiming for asymptotically flat solutions, hereafter we will still leave it explicit in our calculations. 
}
Various constraints on the value of the parameter $\epsilon$ have been obtained from solar and cosmological observations~\cite{Pani:2011mg, Casanellas:2011kf, Avelino:2012ge, Jana:2017ost} to nuclear physics~\cite{Avelino:2012qe, Avelino:2019esh}, including the most stringent bound to date that has been set in consideration of collider experiments~\cite{Delhom:2019wir}. 

Before proceeding to discuss the correspondence between EiBI gravity and GR, the first and foremost task is to determine the deformation matrix $\Omega\indices{^{\mu}_{\nu}}$. 
By applying the definition \eqref{eq:q defn} to the EiBI action \eqref{eq:eibi ac}, we obtain the relation
\be 
\label{eq:eibi q}
q_{\mu\nu} = g_{\mu\nu} + \epsilon \mathcal{R}_{(\mu\nu)} \, ,
\ee
from which it follows that $\Omega\indices{^{\mu}_{\nu}} = \delta\indices{^{\mu}_{\nu}} + \epsilon g^{\mu\alpha} \mathcal{R}_{(\alpha\nu)}$. 
Using this notation, the EiBI Lagrangian density can actually be written in a more compact form as $
\mathcal{L}_G = \big( \sqrt{\detOm} - \lambda \big) / \epsilon$.
With the form \eqref{eq:eibi q} of the connection-compatible metric $q_{\mu\nu}$ worked out, the metric field equations \eqref{eq:RBG eom1} of this theory can then be expressed as
\be 
\label{eq:eibi eom}
\sqrt{-q} \, q^{\mu\nu} = \sqrt{-g} \left( \lambda g^{\mu\nu} - \epsilon T^{\mu\nu} \right) \, . 
\ee
We can now solve for $\Omega\indices{^{\mu}_{\nu}}$ in terms of $g_{\mu\nu}$ and the stress-energy tensor by inserting eq.~\eqref{eq:Om defn} in the metric field equations, yielding
\be 
\label{eq:eibi Om}
[\Omega^{-1}]\indices{^{\mu}_{\nu}} = \frac{1}{\sqrt{\detOm}} \left( \lambda \delta\indices{^{\mu}_{\nu}} - \epsilon T\indices{^{\mu}_{\nu}} \right) \, . 
\ee
For our purposes, it turns out to be more practical to express $\Omega\indices{^{\mu}_{\nu}}$ using quantities in the Einstein frame, which can be achieved with the aid of the mapping relations \eqref{eq:map_L} and \eqref{eq:map_x tildx}. 
We therefore rewrite eq.~\eqref{eq:eibi Om} as 
\be 
\label{eq:eibi Om_Ein}
[\Omega^{-1}]\indices{^{\mu}_{\nu}} = \tilde{f}_1(\tildx, \phi) \, \delta\indices{^{\mu}_{\nu}} + \tilde{f}_2(\tildx, \phi) \, \tildx\indices{^{\mu}_{\nu}} \, ,
\ee
where~\cite{Afonso:2018hyj}
\be 
\tilde{f}_1 = 1 - \frac{\epsilon}{2} \left( \tildx - \tildx \, \del_{\tildx} \tildL \right) \, , \quad \tilde{f}_2 = -\epsilon \, \del_{\tildx} \tildL \, .
\ee

With an eye to generating a solution of EiBI gravity in this work, we focus on the inverse mapping from GR with a minimally coupled free massless scalar field back to EiBI gravity coupled to a scalar field described by the Lagrangian density $\mathcal{L}_m$.
In this case, we start with the scalar Lagrangian density 
\be 
\label{eq:eibi tildL}
\tildL(\tilde{X}, \phi) = \tildx
\ee
in the Einstein frame, and find the corresponding Lagrangian density $\mathcal{L}_m(X, \phi)$ associated with the EiBI theory. 
This can be done by suitably rearranging terms in eq.~\eqref{eq:map_L}, and then making use of eq.~\eqref{eq:map_x tildx} to arrive at 
\be 
\mathcal{L}_m(\tilde{X}, \phi) = 2 \mathcal{L}_G + \sqrt{\detOm} \left( \tildx \del_{\tildx} \tildL - \tildL \right) \, . 
\ee 
Substituting the EiBI Lagrangian and eq.~\eqref{eq:eibi tildL} into the expression above, we get
\be
\mathcal{L}_m(\tildx, \phi) = \frac{2}{\epsilon} \left( \sqrt{\detOm} - \lambda \right) \, .
\ee
For the scalar field model \eqref{eq:eibi tildL} that we are dealing with here, the deformation matrix \eqref{eq:eibi Om_Ein} has the form
\be 
{\color{black}\big[ \Omega^{-1}(\tildx) \big]}\indices{^{\mu}_{\nu}} = \delta\indices{^{\mu}_{\nu}} - \epsilon \tildx\indices{^{\mu}_{\nu}} \, ,
\ee
which results in the relation 
\be 
\label{eq:eibi metric map}
g_{\mu\nu} = q_{\mu\nu} - \epsilon \tildx_{\mu\nu} \, ,
\ee
as well as allowing us to convert between $\tildx$ and $X$ via $\tildx = X / (1 + \epsilon X )$. 
The determinant of $\Omega\indices{^{\mu}_{\nu}}$ can then be computed to give $\abs{\mathbf{\Omega}(X)} = 1 + \epsilon X$, and as a result, we obtain $\mathcal{L}_m$ as a function of the quantities in the EiBI frame~\cite{Afonso:2018hyj}:
\be 
\label{eq:eibi Lm}
\mathcal{L}_m(X, \phi) = \frac{2}{\epsilon} \left( \sqrt{1 + \epsilon X} - \lambda \right) \, ,
\ee
which possesses the square-root structure that is characteristic of Born-Infeld theories of matter~\cite{Born:1934gh}. 
According to \eqref{eq:T}, the scalar field in the EiBI frame has a stress-energy tensor
\be
\label{eq:eibi T}
T_{\mu\nu} = \frac{1}{\sqrt{1 + \epsilon X}} \, X_{\mu\nu} - \frac{\sqrt{1 + \epsilon X} - \lambda}{\epsilon} \, g_{\mu\nu} \, .
\ee
This Born-Infeld type of scalar field has been investigated, for instance, in the context of wormholes~\cite{Lu:2002qja} and cosmological solutions with a late-time accelerating expanding phase~\cite{Lu:2003qg, Fang:2006zr, Jana:2016uvq}. 

\section{An exact rotating solution of EiBI gravity}
\label{sec:new soln}
In this section, we take advantage of the machinery explored so far to construct an exact solution of EiBI gravity by mapping from a known seed metric in GR. 
First, let us continue considering the MES theory that was touched upon in the prior section. 
Its action is given by 
\be 
\mathcal{S} = \frac{1}{2} \int d^4x \sqrt{-g} \Big[ R - (\del_{\mu} \phi)(\del^{\mu} \phi) \Big] \, ,
\ee
and the field equations are 
\be 
\label{eq:MES eom}
R_{\mu\nu} = \del_{\mu} \phi \, \del_{\nu} \phi \, , \qquad \nabla_{\mu}\nabla^{\mu}\phi = 0 \, . 
\ee
Previously, an exact rotating solution of the MES system was obtained in~\cite{Bogush:2020lkp}.\footnote{
In their notation, the scalar field is normalized by a factor of $\sqrt{2}$.
}
The line element of the geometry has the form 
\be 
\label{eq:MES rotsol}
\begin{aligned}
\dd s_{\text{GR}}^2 &= -f(\dd t - \omega \dd \varphi)^2 + f^{-1} \, h_{ij} \dd x^i \dd x^j \, , \\
h_{ij} \dd x^i \dd x^j &= H(r, \theta) (\dd r^2 + \Delta \dd \theta^2) + \Delta \sin^2 \theta \, \dd \varphi^2 \, ,
\end{aligned}
\ee
with the various functions given by 
\be
f(r, \theta) = \frac{\Delta - a^2 \sin^2 \theta}{\rho^2} \, , \quad \omega(r, \theta) = -\frac{2aMr \sin^2 \theta}{\Delta - a^2 \sin^2 \theta} \, , \quad H(r, \theta) = \frac{\Delta - a^2 \sin^2 \theta}{\Delta} \, \zeta(r, \theta) \, , \nn
\ee
where $\Delta(r) = r^2 - 2Mr + a^2$ and $\rho^2(r, \theta) = r^2 + a^2 \cos^2 \theta$.
The effect of the scalar field is encoded in the function
\be
\label{eq:zeta}
\zeta(r, \theta) = \left( 1 + \frac{M^2 - a^2}{\Delta} \, \sin^2 \theta \right)^{-\Sigma^2 / (M^2 - a^2)} \, ,
\ee
where $\Sigma$ is the scalar charge parameter, which, together with the mass $M$ and the spin $a$,  parametrizes the solution. 
For $\Sigma = 0$, \eqref{eq:MES rotsol} reduces to the Kerr line element as expected.
One should note that in the non-rotating limit $a \to 0$, the line element does not boil down to that of FJNW since there is still $\theta$ dependence in $\zeta(r, \theta)$ induced by the scalar field. 
Instead, we recover a particular case of the generalized Zipoy-Vorhees solution ($\gamma$-metric) in the MES theory that was obtained in~\cite{Turimov:2018guy}.\footnote{
To be more specific, the non-rotating limit of \eqref{eq:MES rotsol} corresponds to the scalar-modified $\gamma$-metric given in~\cite{Turimov:2018guy} with $\epsilon = 1$, $\gamma = 1$ and $\gamma_* = \Sigma/M$. 
}
Finally, the solution of the scalar field is~\cite{Bogush:2020lkp}
\be
\label{eq:MES scalar sol}
\phi(r) = \frac{\Sigma}{\sqrt{2(M^2 - a^2)}} \, \log \left( \frac{r - M + \sqrt{M^2 - a^2}}{r - M - \sqrt{M^2 - a^2}} \right) \, , 
\ee
and it can be verified explicitly that the field equations \eqref{eq:MES eom} are indeed satisfied. 
Furthermore, the Ricci scalar of the solution reads
\be 
\label{eq:MES Ricci}
R = \frac{2\Sigma^2}{\rho^2 \Delta} \left( 1 + \frac{M^2 - a^2}{\Delta} \sin^2 \theta \right)^{\Sigma^2/(M^2 - a^2)}.
\ee
We see that for $\Sigma \neq 0$ and $ 0 \leq a/M \leq 1$, besides the ``ring'' singularity (at $r = 0, \, \theta = \pi/2$) inherited from the original Kerr solution, the event horizons located at $r_{\pm} = M \pm \sqrt{M^2 - a^2}$ are now replaced by curvature singularities as well due to the divergent behavior of the scalar field on these hypersurfaces. 
Hence, the solution \eqref{eq:MES rotsol} is well defined only for $r \in (r_+, \infty)$, and it describes a scalar-deformed Kerr spacetime produced by a compact rotating object featuring a naked singularity~\cite{BenAchour:2020fgy}.
As we come close to the naked singularity at $r = r_+$, the Ricci scalar \eqref{eq:MES Ricci} scales as $R \sim \Sigma^2 \rho^{-2} \Delta^{-1-\Sigma^2/(M^2-a^2)}$. 

With everything in place, it is now straightforward to employ the mapping discussed in section~\ref{sec:EiBI map} to find the spacetime solution of EiBI gravity that is in correspondence with the rotating solution \eqref{eq:MES rotsol} of the MES theory.
It was shown there that a free canonical massless scalar field in GR is mapped into a Born-Infeld scalar field with Lagrangian density \eqref{eq:eibi Lm} in the EiBI frame. 
For clarity of illustration, we follow earlier notations and use $q_{\mu\nu}$ to denote the metric associated with the line element \eqref{eq:MES rotsol}. $g_{\mu\nu}$, on the other hand, denotes the corresponding EiBI metric in question. 
Owing to the scalar field being spherically symmetric, the EiBI metric $g_{\mu\nu}$ acquired using the relation \eqref{eq:eibi metric map} differs from the GR metric $q_{\mu\nu}$ only in the $rr$-component:
\be 
\begin{aligned}
g_{\mu\nu} &= q_{\mu\nu} \qquad \text{for $\mu$, $\nu \neq r$ ,} \\
g_{rr} &= \left[ 1 - \epsilon q^{rr} (\del_r \phi)^2 \right] q_{rr} = \frac{\rho^2}{\Delta} \, \zeta(r, \theta) - \frac{2 \epsilon \Sigma^2}{\Delta^2} \, . 
\end{aligned}
\ee
Consequently, the EiBI counterpart of the rotating MES solution \eqref{eq:MES rotsol} takes the form
\be 
\begin{aligned}
\label{eq:eibi rotsol}
\dd s_{\text{EiBI}}^2 = 
&- \left( 1 - \frac{2Mr}{\rho^2} \right) \dd t^2 - \frac{4Mar \sin^2 \theta}{\rho^2} \, \dd t \, \dd \varphi + \left[ \frac{\rho^2}{\Delta} \, \zeta(r, \theta) - \frac{2 \epsilon \Sigma^2}{\Delta^2} \right] \dd r^2 \\
&+ \rho^2 \zeta(r, \theta) \, \dd \theta^2 + \left( r^2 + a^2 + \frac{2 M a^2 r}{\rho^2} \, \sin^2 \theta \right) \sin^2 \theta \, \dd \varphi^2 \, .
\end{aligned}
\ee 
Once again, one can check explicitly that the field equations \eqref{eq:eibi eom} of EiBI gravity with a Born-Infeld scalar source \eqref{eq:eibi Lm} are satisfied by the solution \eqref{eq:eibi rotsol}.  
The mapping \eqref{eq:Om defn} that was applied did not involve any manipulation of the matter fields at all, meaning that the scalar fields on both sides of the correspondence have identical configurations \eqref{eq:MES scalar sol}.
Note that for $\Sigma = 0$, \eqref{eq:eibi rotsol} reduces to the Kerr line element. 
This is consistent with the property of RBGs that their deviations from GR are entirely attributed to the matter sector, and are thus absent in vacuum. 
Likewise, \eqref{eq:eibi rotsol} asymptotically approaches the Kerr solution at large $r$ as well due to the vanishing of the scalar field, which then guarantees that the metric satisfies weak-field tests. 
Analysis with regard to the strong-field properties of the spacetime will be performed afterwards by studying its shadow. 

\section{Properties of the solution}
\label{sec:properties}
\subsection{Curvature and singularities}
\begin{figure}[t]
\centering
\includegraphics[scale=0.5]{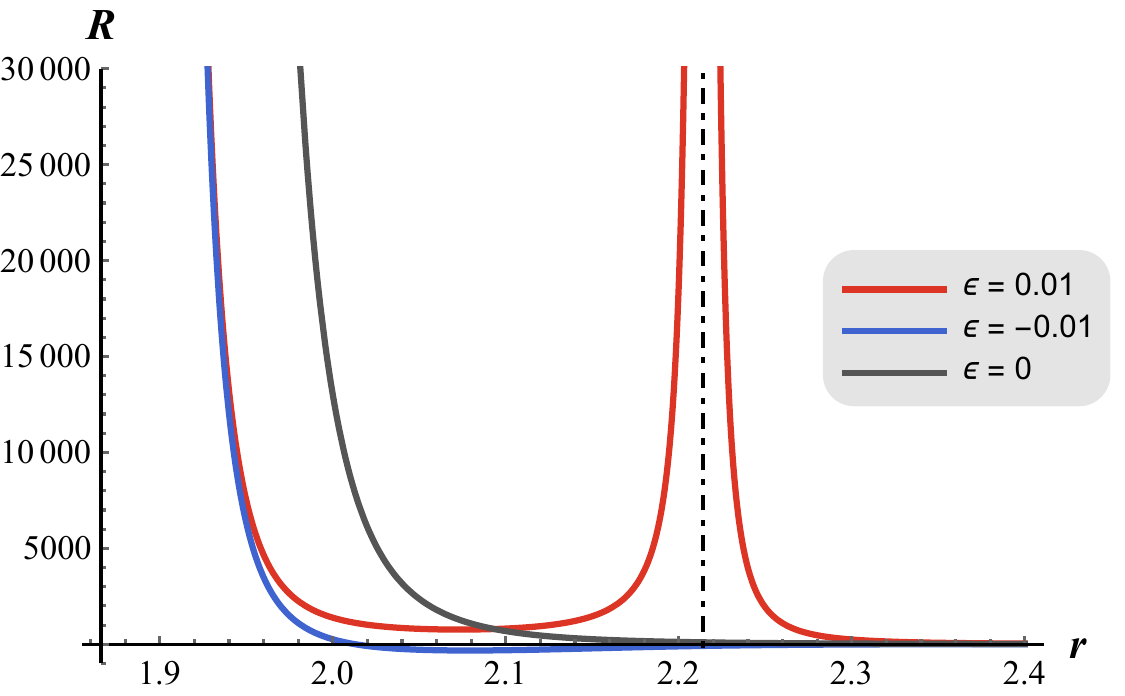}
\caption{Radial profile ($r > r_+$) of the Ricci scalar on the equatorial plane for different values of $\epsilon$, with $\epsilon = 0$ corresponding to the curvature \eqref{eq:MES Ricci} of the rotating MES solution.  The radial distance is in units of $GM$, which is set equal to one, and we have chosen the scalar charge to be $\abs{\Sigma} = 2$. The vertical line marks the value of $r$ at which the additional curvature singularity appears.}
\label{fig:ricci plot}
\end{figure}

We embark on the analysis of the EiBI solution \eqref{eq:eibi rotsol} by inspecting the curvature. Explicit computation of the Ricci scalar $R(g) = g^{\mu\nu} R_{\mu\nu}(g)$ shows that it diverges on the hypersurfaces $\Delta = 0$ and $\rho^2 \zeta \Delta = 2 \epsilon \Sigma^2$, with the former case diverging as $R \sim \rho^{-2} \Delta^{-\Sigma^2/(M^2-a^2)}$ and the latter as $R \sim \epsilon\Sigma^2/(\rho^2 \zeta \Delta - 2 \epsilon \Sigma^2)^2$. 
A closer evaluation makes it clear that the curvature singularities are caused by the trace of the stress-energy tensor \eqref{eq:eibi T} of the Born-Infeld scalar field being divergent: 
\be 
T = \frac{2\Sigma^2 - 4 \rho^2 \zeta \Delta/\epsilon}{\sqrt{\rho^2 \zeta \Delta(\rho^2 \zeta \Delta - 2\epsilon\Sigma^2)}}\Bigg\rvert_{\Delta \text{ or } F \to 0} + \frac{4\lambda}{\epsilon} \to \infty \, , 
\ee
where we have defined $F(r, \theta) \equiv \zeta - 2 \epsilon \Sigma^2 / (\rho^2 \Delta)$. 
We see that the curvature singularity of the MES solution \eqref{eq:MES rotsol} at the hypersurface $r = r_+$ on which $\Delta = 0$ is still present after the mapping.\footnote{
Since $r = r_+$ is already a curvature singularity, we will only focus on regions exterior to it and ignore the interior structure at, for example, $r = r_- = M^2 - \sqrt{M^2-a^2}$. 
} 
Nevertheless, unlike in GR where the divergence of the stress-energy tensor is directly transferred to the curvature through the Einstein equations, in EiBI gravity, instead, the matter and geometric sectors couple in a different manner (cf.~\eqref{eq:eibi eom}) such that the curvature divergence at $r = r_+$ gets ``softened'' by roughly a factor of $\Delta$ compared to that of \eqref{eq:MES Ricci}. 
The Ricci scalar as a function of $r$ on the equatorial plane ($\theta = \pi / 2$) is depicted in figure~\ref{fig:ricci plot}, which captures the major traits of its behavior. 
Another divergence of the Ricci scalar occurs when $F(r, \theta) = 0$, i.e. $\rho^2 \zeta \Delta = 2 \epsilon \Sigma^2$. 
For the EiBI parameter $\epsilon > 0$, the right-hand side is positive, and we can infer that the condition will be satisfied at a radial coordinate greater than $r_+$. This explains why there is an additional curvature singularity exterior to the previous one, as is apparent from figure~\ref{fig:ricci plot}.
For $\epsilon < 0$, the right-hand side becomes negative, and the condition can only be met within $r < r_+$. 
In this case, the additional curvature singularity, if it exists, is covered by the previous singularity at $r = r_+$ and does not concern us. 

The causal structure of the $r = r_+$ hypersurface is determined by the norm of its normal vector field $\underline{n} \eqdef \del_r$, which is given by 
\be 
n_{\mu} n^{\mu} = g^{rr} = \frac{\Delta^2}{\rho^2 \zeta \Delta - 2\epsilon\Sigma^2} \, .
\ee
The expression vanishes exactly when $\Delta = 0$, thus $r = r_+$ is a null singularity.
In addition, it turns out that the Killing vector of the combination $\underline{K} = \del_t + \left[ a/(r^2+a^2) \right]\del_{\varphi}$ becomes null on this hypersurface as well~\cite{Carroll:2004st}, where $\del_t$ and $\del_{\varphi}$ are the Killing vectors associated with the stationarity and the axial symmetry of the metric \eqref{eq:eibi rotsol}. 
As a result, the hypersurface $r = r_+$ is also a Killing horizon, just as in the case of its GR counterpart \eqref{eq:MES rotsol}. 

The other singular hypersurface $F(r, \theta) = 0$ (or $\rho^2 \zeta \Delta = 2 \epsilon\Sigma^2$) has a normal vector $\underline{N}$ with norm 
\be
N_{\mu} N^{\mu} = g^{rr} \left( \pdv{F}{r} \right)^2 + g^{\theta\theta} \left( \pdv{F}{\theta} \right)^2 \, .
\ee
To examine the causal structure of this hypersurface on which $g_{rr}$ vanishes, it is evident that our current choice of coordinates is not adequate and has to be altered to avoid an ill-defined norm of $\underline{N}$. 
The idea is to perform a coordinate transformation in order to remove the coordinate singularity caused by the problematic $rr$-component of the metric. 
To begin with, we find the (normalized) principal null vectors $l_{\pm}^{\mu}$ of the spacetime \eqref{eq:eibi rotsol}, which are given by
\be
l_{\pm}^{\mu} \del_{\mu} = \frac{r^2+a^2}{\Delta} \, \del_t \pm \frac{1}{\sqrt{F(r, \theta)}} \, \del_r + \frac{a}{\Delta} \, \del_{\varphi} \, .
\ee
Paralleling the techniques of bringing the Kerr metric into the Eddington-Finkelstein form, we take the ingoing direction $l_-^{\mu}$ and write
\be
\label{eq:sing EF}
\dd v \eqdef \dd t + \frac{r^2+a^2}{\Delta} \, \dd x \, , \qquad \dd \bar{\varphi} \eqdef \dd \varphi + \frac{a}{\Delta} \, \dd x \, ,
\ee
where $v$ and $\bar{\varphi}$ are two new coordinates, and $x$ is ``defined'' by $\dd x \eqdef \sqrt{F(r, \theta)} \, \dd r$. 
So far, the relations in \eqref{eq:sing EF} do not constitute a valid coordinate transformation due to the dependence on $\theta$ through $F(r, \theta)$ (see~\cite{AzregAinou:2011fq} for a similar situation).  
Fortunately, we have no intention of carrying out further analysis based on the transformed line element except to extract the causal character of the $F(r, \theta) = 0$ singular hypersurface. 
Since there is no signature change in the metric on this hypersurface ($g_{rr} = 0$, $g_{\theta\theta} > 0$), it suffices to compute the norm of the $F(r, \pi / 2) = 0$ or $F(r, 0) = 0$ submanifold on which $\pdv*{F}{\theta} = 0$ and the coordinate transformation \eqref{eq:sing EF} is well defined. 
The induced metric on the equatorial plane can then be rewritten in an Eddington-Finkelstein form such that the line element becomes
\be 
\label{eq:sing EFmet}
\dd s_{\text{eq}}^2 = 
- \left( 1 - \frac{2M}{r} \right) \dd v^2 + 2 \, \dd v \dd x - \frac{4Ma}{r} \, \dd v \dd \bar{\varphi} - 2a \, \dd x \dd \bar{\varphi} + \left( r^2 + a^2 + \frac{2 M a^2}{r} \right) \dd \bar{\varphi}^2 \, .
\ee 
From our ``definition'' of $x$, we see that the surface satisfying $F(r, \pi / 2) = 0$ belongs to the family of $x = \text{constant}$ hypersurfaces on the equatorial plane. 
The vector field normal to these hypersurfaces has norm
\be 
\bar{N}_{\mu} \bar{N}^{\mu} = \frac{\Delta}{r^2} 
\ee
at $\theta = \pi / 2$ according to \eqref{eq:sing EFmet}, which is clearly positive for $r > r_+$. 
The same conclusion can be drawn by considering the $F(r, 0) = 0$ submanifold. 
Therefore, for $\epsilon > 0$ the hypersurface $\rho^2 \zeta \Delta = 2 \epsilon\Sigma^2$ is in fact a timelike singularity. 

\subsection{Shadow}
\label{sec:sha}
Now let us turn our attention to the shadow cast by the EiBI rotating compact scalar object. 
The vital ingredient in determining the shadow is the surrounding photon region in which photons undergo unstable spherical motions around the central object. 
The null geodesic equations of the spacetime \eqref{eq:eibi rotsol} can be derived using the Hamilton-Jacobi approach. With the absence of an analog of the Carter constant for this spacetime, the geodesic equations for the radial and polar angular components are in general not decoupled; however, we can render them so by only considering photon trajectories close to the equatorial plane and work with $\theta = \pi / 2 + \delta \theta$. 
The majority of the orbits are certainly not confined to the equatorial plane. 
That said, such an approximation works well enough when the observer is located on the equatorial plane far away from the object where photons essentially arrive at a polar angle $\theta \approx \pi / 2$. 
This scheme was considered in~\cite{Abdujabbarov:2016hnw} when studying the shadow of a rotating black hole whose metric functions have rather complicated forms. 
In this work, we expect that the analytic results obtained in this way are still able to shed light on some key features regarding how the shadow is affected by the presence of the parameters in the model, without needing to resort to a full-blown ray-tracing analysis.

On the equatorial plane, $\rho^2 = r^2$, and the geodesic equations for the $t$ and $\varphi$ components read
\begin{align}
r^2 \dot{t} &= a(L_z - a E) + \frac{r^2 + a^2}{\Delta} \left[ E(r^2 + a^2) - a L_z \right] \, , \\
\label{eq:geo phi}
r^2 \dot{\varphi} &= L_z - aE + \frac{a}{\Delta} \left[ E(r^2 + a^2) - a L_z \right] \, ,
\end{align}
where the dot denotes the derivative with respect to an affine parameter $\tau$, and the energy $E$ and azimuthal angular momentum $L_z$ of the particle are conserved quantities associated with the Killing vector fields $\del_t$ and $\del_{\varphi}$, respectively. 
These two equations have the same forms as those of a Kerr black hole, as the metric components $g_{tt}$, $g_{t\varphi}$, and $g_{\varphi\varphi}$ in \eqref{eq:eibi rotsol} are free of Born-Infeld and scalar corrections.
The corrections come into play in the equations for the $r$ and $\theta$ components, which are given by
\begin{align}
\label{eq:geo r}
\left( \frac{r^2}{E} \right)^2 \dot{r}^2 &= \left( \zetaeq - \frac{2\epsilon\Sigma^2}{r^2 \Delta} \right)^{-1} \mathfrak{R}(r) \, , \\
\label{eq:geo theta}
\left( \frac{r^2}{E} \right)^2 \dot{\delta\theta}^2 &= \frac{1}{\zetaeq^2} \, \eta \, ,
\end{align}
where 
\be 
\mathfrak{R}(r) = (r^2 + a^2 - a \xi)^2 - \Delta \left[ \frac{\eta}{\zetaeq} + (\xi - a)^2 \right] \, ,
\ee
and $\zetaeq(r)$ is the function \eqref{eq:zeta} evaluated on the equatorial plane $\theta = \pi / 2$, i.e. 
\be 
\zetaeq(r) \equiv \zeta(r, \pi/2) = \left[ \frac{(r-M)^2}{\Delta} \right]^{-\Sigma^2/(M^2-a^2)} \, .
\ee
We have introduced two dimensionless parameters $\xi \equiv L_z/E$ and $\eta \equiv \mathcal{Q}/E^2$, with $\mathcal{Q}$ being the Carter-like constant for orbits near the equatorial plane. 
Note that eqs.~\eqref{eq:geo r} and \eqref{eq:geo theta} make sense only when $\zetaeq - 2 \epsilon \Sigma^2 / (r^2 \Delta) > 0$ and $\zetaeq(r) \neq 0$. 
Since the spacetime \eqref{eq:eibi rotsol} is only well defined outside the naked singularity at $r = r_+$, $\zetaeq(r)$ never reaches zero, and thus the second requirement does not trouble us in any way. 
As argued earlier, the first requirement is automatically satisfied for $r > r_+$ when $\epsilon < 0$, whereas when $\epsilon > 0$, it further restricts the discussion of geodesics specifically to regions exterior to the additional timelike singularity.
We will come back to this point later as a consistency check. 

Equations~\eqref{eq:geo r} and \eqref{eq:geo theta} govern the propagation of light near the equatorial plane of the spacetime \eqref{eq:eibi rotsol}. 
We are particularly interested in unstable spherical lightlike orbits that constitute the photon region.
A photon orbit of constant Boyer-Lindquist radius $\rph$ is characterized by $\dot{r}\rvert_{\rph} = 0$ and $\ddot{r}\rvert_{\rph} = 0$, which in turn requires $\mathfrak{R}(\rph) = 0$ and $\mathfrak{R}'(r)\rvert_{\rph} = 0$, where the prime stands for the derivative with respect to $r$. 
These conditions demand that for spherical photon orbits the constants of motion must satisfy
\begin{align}
\label{eq:shadow xi}
\xi(\rph) &= \eval{\frac{2Ma(\ycal-r) + r\sqrt{a^2(r - 2\ycal)^2 + r(r - 4\ycal) \big[ r^2 + 2M(\ycal - r) - 2\ycal r \big]}}{r^2 + 2M(\ycal - r) - 2\ycal r}}_{r = \rph} \, , \\
\label{eq:shadow eta}
\eta(\rph) &= \zetaeq \left[ \frac{1}{\Delta} \, (r^2 + a^2 - a \xi)^2 - (\xi - a)^2 \right]\Bigg\rvert_{r = \rph} \, , 
\end{align}
where
\be
\mathcal{Y}(r) \equiv \frac{\Delta}{\zetaeq(\Delta/\zetaeq)'} = \frac{1}{2}\frac{(r-M)\Delta}{(r-M)^2-\Sigma^2} \, .
\ee
The kinematic quantities $\xi$ and $\eta$ of these photon orbits are parametrized solely by the orbital radius $\rph$, the range of which is dictated by the condition $\eta(\rph) \geq 0$~\cite{Teo:2003}. 
Another significant property of the orbits in the photon region is that they are unstable under radial perturbations, i.e. $\mathfrak{R}''(r)\rvert_{\rph} > 0$. 
That way, when light originating from a distant source travels near these unstable orbits, it will either spiral in and hit the naked singularity, or escape and reach an observer at infinity, depending on the impact parameter. 
The photon region thus defines a bright ring on the boundary of an unilluminated shadow region in the sky of the observer.

\begin{figure}[t]
\centering
\includegraphics[scale=0.45]{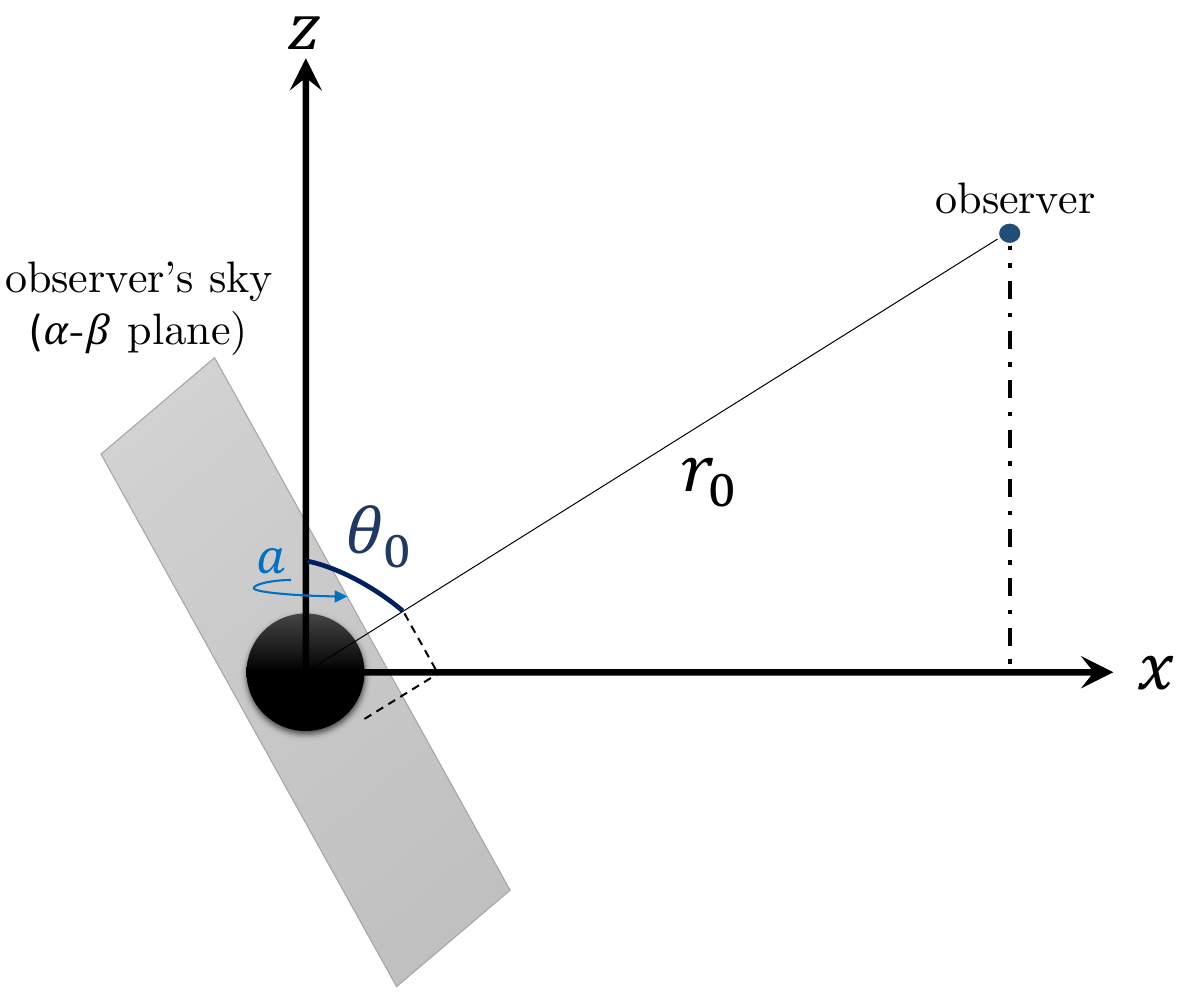}
\caption{Schematic illustration of the celestial coordinates. Due to the azimuthal symmetry of the object of interest, which is situated at the origin of the coordinate system, we can choose the observer to be on the $x$-$z$ plane, and the symmetry axis of the object to be aligned with the $z$-axis. The observer sees the shadow of the object as being projected onto the $\alpha$-$\beta$ plane, which is normal to the direction from the object to the observer.}
\label{fig:celestial}
\end{figure}

Given that the spacetime \eqref{eq:eibi rotsol} is asymptotically flat, we can set up a coordinate system as in figure~\ref{fig:celestial}, and adopt the celestial coordinates $(\alpha, \beta)$ in the observer's sky to describe the outline of the shadow. 
The coordinates $\alpha$ and $\beta$ represent the apparent perpendicular distances from the boundary of the shadow to the symmetry axis of the object and to the equatorial plane, respectively. 
For an observer at a distance $r_0$ far away from the object, it can be shown that \cite{Vazquez:2003zm}
\be
\alpha = \lim_{r_0 \to \infty} \left( -r_0^2 \sin \theta_0 \, \frac{d\varphi}{dr}\Bigg\rvert_{r_0, \, \theta_0} \right) \, , \qquad \beta = \lim_{r_0 \to \infty} \left( r_0^2 \, \frac{d (\delta\theta)}{dr} \Bigg\rvert_{r_0, \,  \theta_0} \right) \, ,
\ee
where $\theta_0$ is the inclination angle between the symmetry axis of the object and the direction of the observer (see figure~\ref{fig:celestial}). 
In our case, we are limited to light rays propagating near the equatorial plane, so we shall take $\theta_0 = \pi / 2$. 
Now, $d \varphi / dr$ and $d (\delta \theta) / dr$ can be obtained by combining the geodesic equations \eqref{eq:geo phi}--\eqref{eq:geo theta}.
Substituting them into the expressions above and then retaining only the leading terms in $r_0$ yields
\be 
\alpha = -\xi \, , \qquad \beta = \pm \sqrt{\eta} \, .
\ee
That is, the position of the image of the shadow contour on the celestial plane is fully determined by
eqs.~\eqref{eq:shadow xi} and \eqref{eq:shadow eta} for unstable spherical photon orbits.
Notice that the Born-Infeld scale $\epsilon$ does not enter into these parameters, signalling that the appearance of the shadow will not be affected by the higher-order derivative terms, as well as the length scale at which they manifest, in \eqref{eq:eibi ac} and \eqref{eq:eibi Lm}.
Therefore, we are in fact simultaneously studying the shadow image of both the MES solution \eqref{eq:MES rotsol} and the new EiBI solution \eqref{eq:eibi rotsol}. 
Such a degeneracy can be accounted for by the fact that the mapping relation \eqref{eq:Om defn} will leave the photon region unchanged so long as the null geodesic equations are separable and the field configurations for (bosonic) matter are spherically symmetric.\footnote{
Recall that we were able to work with separable null geodesic equations \eqref{eq:geo r} and \eqref{eq:geo theta} since we had \emph{forced} them to be so by fixing the polar angle $\theta \approx \pi /2$. 
}
The intrigued reader is referred to appendix~\ref{app:equi shadow} for an explicit proof.
This degeneracy applies to other solutions generated through the RBG/GR mapping involving spherically symmetric matter sources in several preceding works, e.g.~\cite{Afonso:2019fzv}, and is just one of many instances~\cite{Vincent:2015xta, Cunha:2018gql, Abdikamalov:2019ztb} in which the shadow displays its insensitivity to deformations of the metric.  

\begin{figure}[t]
\centering  
\subfloat{\includegraphics[width=0.33\columnwidth]{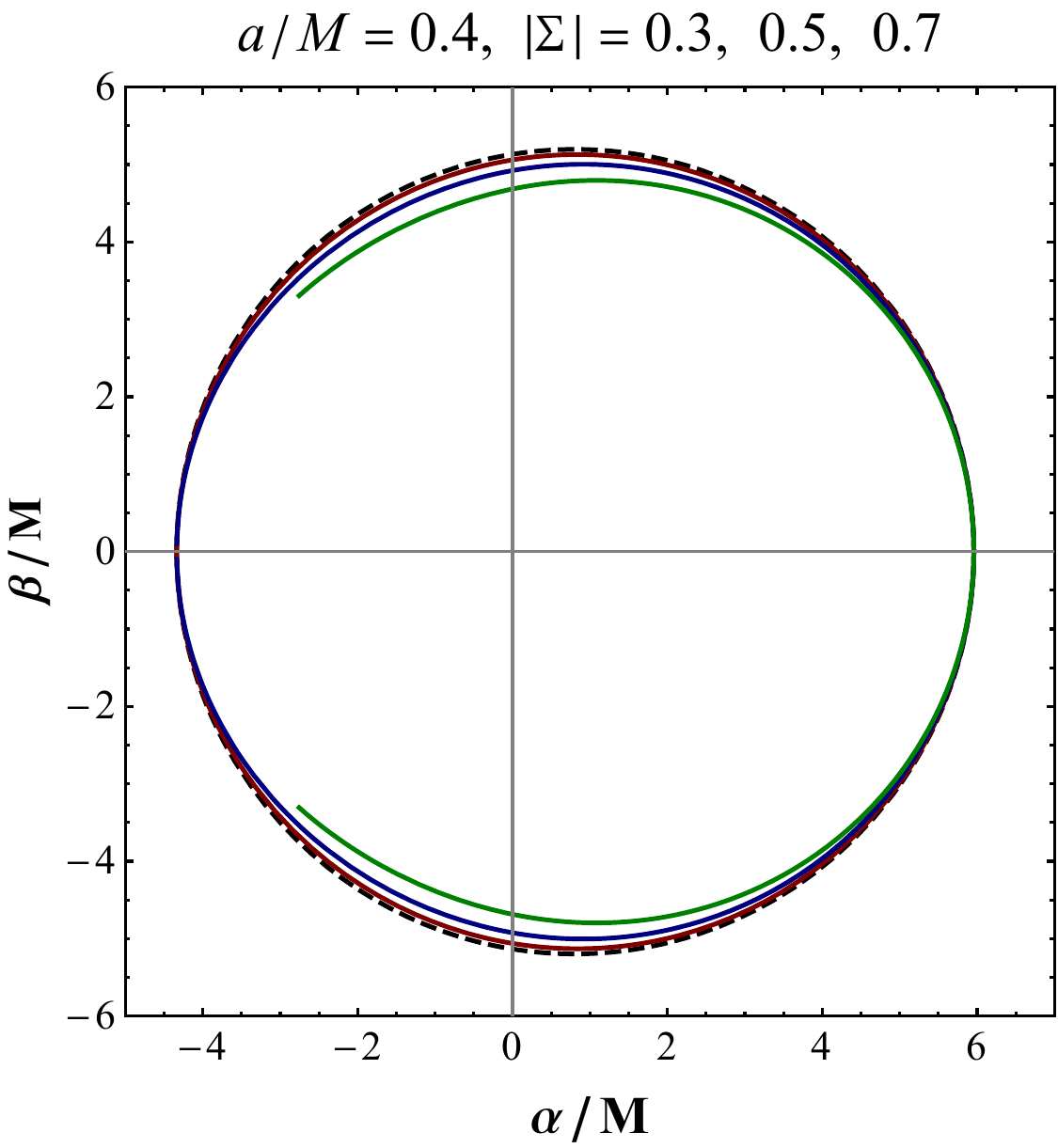}}\hfill%
\subfloat{\includegraphics[width=0.33\columnwidth]{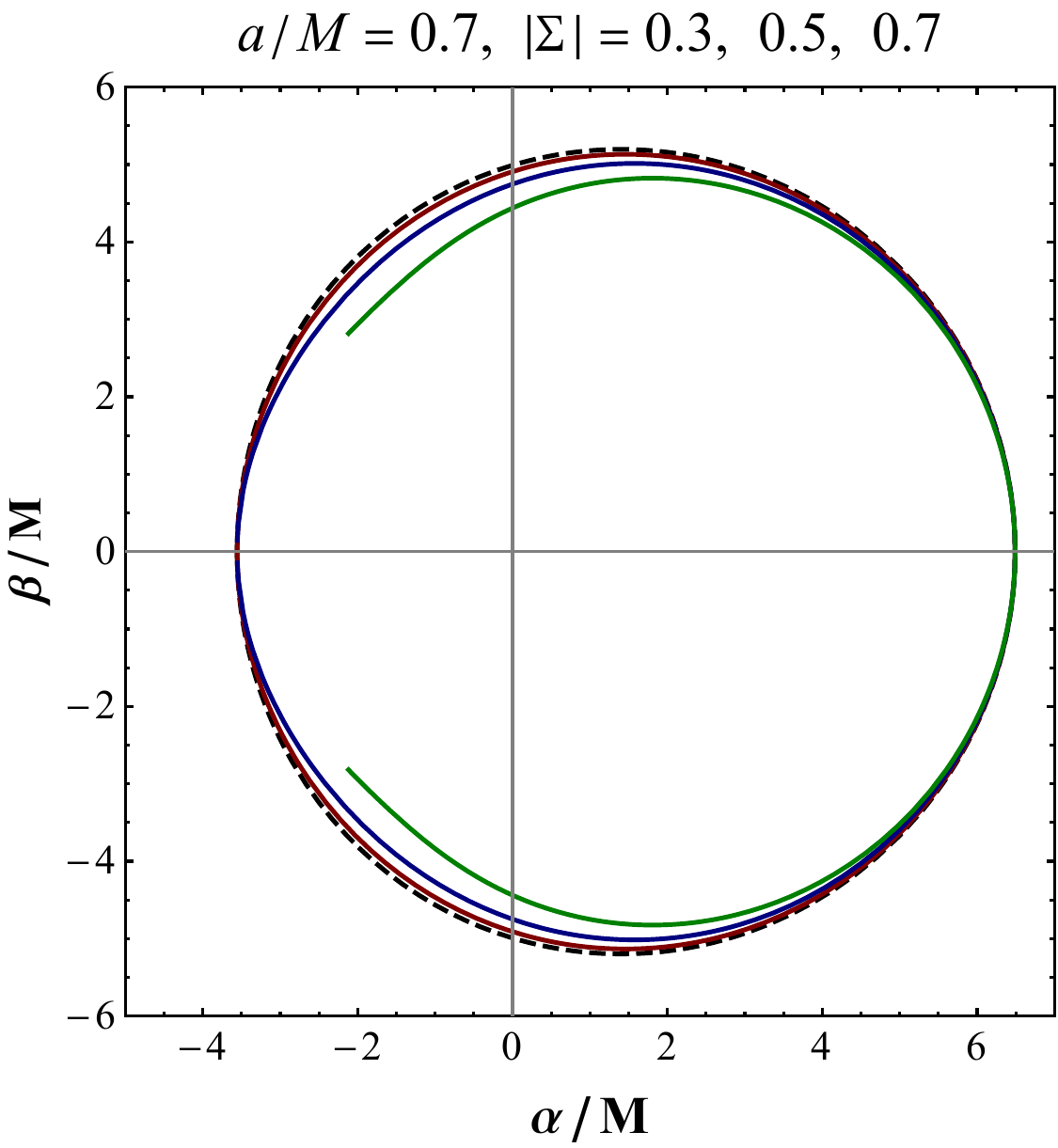}}\hfill%
\subfloat{\includegraphics[width=0.33\columnwidth]{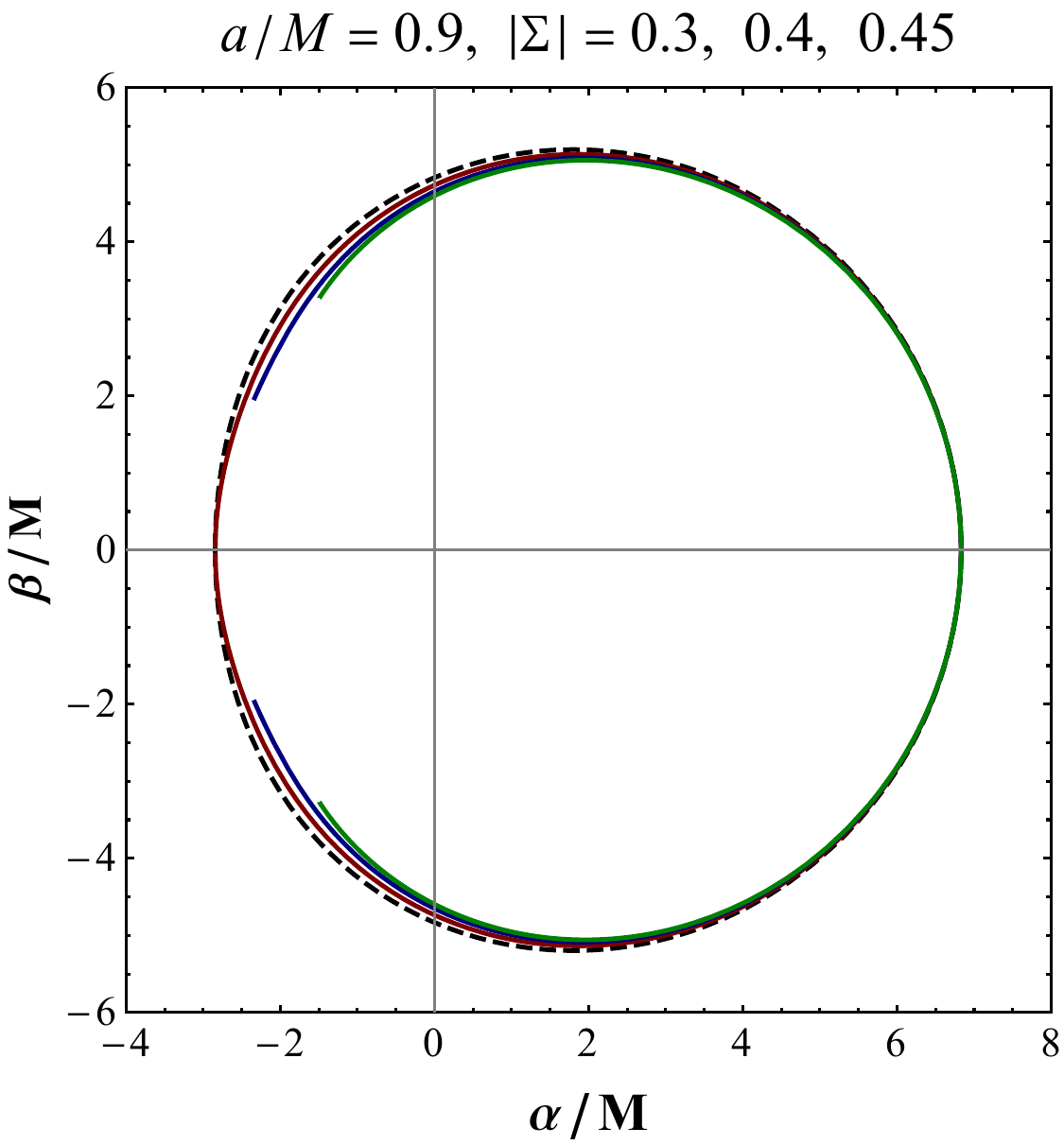}}%
\caption{Shadow cast by both naked singularities \eqref{eq:MES rotsol} and \eqref{eq:eibi rotsol} as seen by a distant observer on the equatorial plane for different values of $a/M$ and $\abs{\Sigma}$. The dashed black curve illustrates the shadow contour of a Kerr black hole ($\Sigma=0$), whereas the red, blue, and green curves correspond to the cases where $\abs{\Sigma}$ takes the values above each panel sequentially from left to right.
}
\label{fig:shadow}
\end{figure}

Figure~\ref{fig:shadow} offers a visualization of the apparent shape of \emph{both} the MES and the EiBI rotating scalar objects. 
We can see from the figure that, despite the underlying entities being naked singularities, they are still capable of casting shadows thanks to the existence of an external photon region. 
That the shadow is not a signature unique to black holes has long been known (see, e.g.,~\cite{Kumar:2020oqp}), and it has been established that naked singularities are also candidates that can produce them~\cite{Nakao:2002kc, Bambi:2008jg, Ortiz:2015rma, Shaikh:2018lcc}, even when lacking a photon sphere~\cite{Dey:2020haf, Joshi:2020tlq, Dey:2020bgo}.\footnote{
This type of naked singularity is dubbed a strongly naked singularity (SNS)~\cite{Virbhadra:2002ju}, as opposed to the ones that we are considering in this paper, which fall into the category of weakly naked singularity (WNS) since they are covered by photon regions. WNSs can in general give rise to observational features similar to that of black holes~\cite{Virbhadra:2002ju}, and can thus possibly act as black hole mimickers. 
} 
Still, in our case the structure of the shadow will be influenced by the scalar charge parameter $\Sigma$.
In figure~\ref{fig:shadow}, it is clear that with the rotation parameter $a$ fixed, the boundary of the shadow region is closed for small enough $\abs{\Sigma}$, but reduces to an open arc once $\abs{\Sigma}$ exceeds a certain critical value $\Sigmacr$. 
To our surprise, for values of $\abs{\Sigma}$ that lead to closed shadow contours (including the Kerr case $\Sigma = 0$), the boundaries of their corresponding photon regions intersect the equatorial plane in circles of the exact same radii.  
In other words, what these photon regions have in common are the prograde and retrograde circular orbits on the equatorial plane, which is why the left and right endpoints of the closed shadow contours in figure~\ref{fig:shadow} coincide. 
Therefore, as far as the observer on the equatorial plane is concerned, the image of a closed shadow contour nearby the $\alpha$-axis bears great resemblance to that of a Kerr black hole, not to mention that this part of the image is accurate under the equatorial plane approximation $\theta \approx \pi / 2 + \delta\theta$.
A precise description of the overall shape of the shadow as seen from other inclination angles is beyond the scope of the approximation, so we will not dwell too much on it, other than to make a few qualitative comments below.\footnote{
A more complete treatment incorporating numerical ray tracing followed by a quantitative analysis of the size and distortion of the shadow will be carried out elsewhere.}
To an observer on the equatorial plane, besides the fact that the scalar objects and the Kerr black hole have similar shadow sizes, notice from figure~\ref{fig:shadow} that the shadow of the object is more or less deformed vertically when the object picks up a scalar charge. 
The reason is that $\Sigma$ takes part in the metric in a way that it tends to suppress the polar angular speed $\eta$ of the photon passing through the equatorial plane (cf.~\eqref{eq:shadow xi} and \eqref{eq:shadow eta}), which then has a notable effect on the shadow in terms of its ``oblateness''.
This is to be contrasted with known examples in the past where the shadow usually receives a variation in size as well as a horizontal distortion on the left due to parameters such as the electric charge~\cite{2000CQGra..17..123D, 8209515}, the tidal charge~\cite{Amarilla:2011fx}, the NUT charge~\cite{Abdujabbarov:2012bn}, or others in various extensions of the Kerr solution~\cite{Johannsen:2015qca, Ghasemi-Nodehi:2015raa, Younsi:2016azx, Wang:2017hjl, Medeiros:2019cde, Carson:2020dez, Chen:2020aix}. 

\begin{figure}[t]
\centering
\includegraphics[scale=0.5]{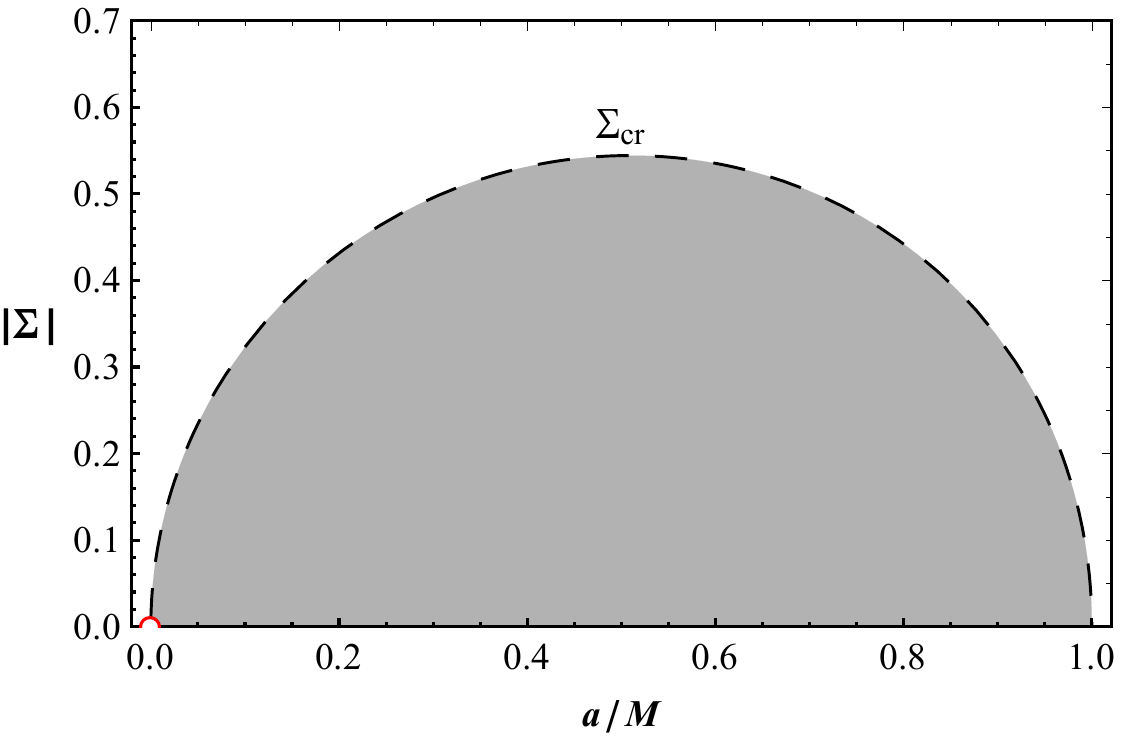}
\caption{Parameter space $\left( a/M, \abs{\Sigma} \right)$ of the naked singularity spacetimes \eqref{eq:MES rotsol} and \eqref{eq:eibi rotsol}. The dashed line indicates the critical scalar charge $\Sigmacr$ for each spin $a$, while parameters lying in the shaded region give rise to closed shadow contours when observed from the equatorial plane. For $a = 0$, there is no critical scalar charge since the circular orbit on the equatorial plane exists for all $\Sigma$.}
\label{fig:scalarbound}
\end{figure}

When $\abs{\Sigma}$ is just slightly above the critical value $\Sigmacr$, we find that the prograde circular photon orbit on the equatorial plane and a few of its neighboring spherical photon orbits with small $\eta$ cease to exist.
For this reason, a tiny segment of the shadow contour near the left endpoint is now missing, and the boundary of the shadow is no longer closed. 
As we further enlarge $\abs{\Sigma}$, the photon region gets elongated radially inwards, but at the same time, an increasing number of prograde spherical photon orbits fail to remain, thus causing the shadow contour to be chipped away from the left (see figure~\ref{fig:shadow}). 
Hence, the scalar charge $\abs{\Sigma}$ has to be below the critical value $\Sigmacr$ so that the prograde circular photon orbit can exist, which is a necessary condition for the shadow contour to be closed. 
The range of the scalar charge that allows a closed shadow contour to occur is shown in figure~\ref{fig:scalarbound}. 
Seeing that the scalar charge is constrained to be small ($\abs{\Sigma} < 1$), we can naturally expect that the corresponding closed shadow contour \emph{as a whole} will deviate only a little from that of a Kerr black hole.
In particular, they will be nearly identical in both the extremal $a \to M$ and non-rotating $a \to 0$ limits, since the closedness requirement alone forces the scalar charge to be practically zero.

The point $a = 0$, which was purposely excluded from the parameter space in figure~\ref{fig:scalarbound}, is a special case that deserves more scrutiny. 
When the spin parameter is exactly zero, it turns out that the only consistent spherical photon orbit that we can work with premised on the assumption $\theta \approx \pi / 2 + \delta\theta$ is the equatorial circular orbit at $r = 3M$ (can be either prograde or retrograde).
This orbit happens to exist for all values of $\Sigma$, so the scalar charge is unconstrained for $a = 0$ within the equatorial plane approximation. 
The information of just the equatorial photon orbit is not yet enough to provide insight into the shadow of the non-rotating case.
The single piece of knowledge of the shadow that we can deduce at this stage is that its left and right endpoints match with those of the shadow produced by a Schwarzschild black hole when seen from the equatorial plane. 

Last but not least, for scalar charge in the range $\abs{\Sigma} \leq \Sigmacr$, it can be confirmed that the stability condition $\mathfrak{R}''(r)\rvert_{\rph} > 0$ holds true for all of the spherical photon orbits characterized by eqs.~\eqref{eq:shadow xi} and \eqref{eq:shadow eta}.
Moreover, one can verify that these orbits satisfy $\eval{[\rho^2 \zeta \Delta]}_{\rph} - 2 \epsilon\Sigma^2 > 0$ as long as $\epsilon \lesssim 0.1$, ensuring that the entire photon region is safe from the additional timelike singularity that is present when $\epsilon > 0$. 

\section{Conclusions and outlook}
\label{sec:conclusion}
Exact rotating solutions of either GR or modified gravity theories in the presence of matter fields are of pivotal importance due to their ability to describe realistic compact objects and to serve as toy models for studying the interplay of strong gravity with matter. 
However, these rotating solutions are currently still scarce, and this may severely hinder the progress of illuminating possible signs of new physics at astrophysical scales.
The simplest of such theories is the minimal Einstein-scalar (MES) theory which exhibits extra long-range effects from a massless scalar field. 
A rotating solution of this theory was discovered recently~\cite{Bogush:2020lkp}, and it was found to possess a naked singularity. 

In an attempt to achieve a solution free from naked singularity, we turned to study the extension of the rotating MES solution in EiBI gravity, a theory that is known for its capability to smooth out singularities by modifications at the Born-Infeld scale $\abs{\epsilon}$. 
It was shown that a theory of EiBI gravity coupled to a matter source admits an Einstein frame representation where the field equations correspond formally to the Einstein equations for an auxiliary metric sourced by the same matter but described by a different matter Lagrangian.
This correspondence provides a mapping between the spacetime solutions of GR and EiBI gravity such that the solution on one side can be obtained from its counterpart on the other side simply through a transformation of the metric. 
In particular, starting with the rotating MES solution as the seed metric, we generated an exact analytic rotating solution of EiBI gravity coupled to a Born-Infeld-type scalar field via the correspondence, where $\epsilon$ and the scalar charge $\Sigma$ parametrize the solution along with the mass and spin $a$. 
Unfortunately, the naked singularity in the MES solution is neither resolved nor is it concealed behind a horizon after mapping the solution into the Born-Infeld setting.
The singularity still resides at what would be the event horizon hypersurface of the new spacetime, only with its curvature divergence suppressed. 
Furthermore, for $\epsilon > 0$ there is in fact an additional timelike singularity exterior to the previous null singularity.  
On top of all that, one can verify using eq.~\eqref{eq:geo r} that radial null rays propagating on the equatorial plane reach both of these singularities in a finite amount of affine time, beyond which the geodesics cannot be further extended. 
This is sufficient to lead us to the conclusion that the naked singularities one may encounter in this new solution are geodesically incomplete. 

Equipped with the solution, we then revealed that there exists a photon region covering the naked singularity without coming in contact with it.
This set the stage for us to proceed to draw connections with observations by analysing the shadow cast by the object. 
In this work, we have limited ourselves to considering a stationary observer on the equatorial plane far away from the object where light rays that reach can be approximated as propagating at a polar angle $\theta \approx \pi / 2$. 
For such an observer, one would merely need the unstable spherical photon orbits near the equatorial plane to describe the approximate shape of the shadow with decent accuracy. 
The calculations were thus considerably simplified, and we were able to obtain immediate analytic results that capture a few prominent features of the shadow.
First, we showed that under circumstances where the null geodesic equations are separable and the matter fields are spherically symmetric, the shadows are completely identical on both sides of the mapping, independent of the high-energy Born-Infeld corrections to GR and to the matter sector. 
So, we have essentially studied the approximate shadow corresponding to both the seed MES solution and the generated EiBI solution at the same time.  
Next, we found that the prograde spherical photon orbits that remain on or close to the equatorial plane begin to disappear as the scalar charge exceeds a critical value that depends on the spin parameter. 
Without these unstable orbits, the shadow boundary will no longer be \emph{closed} as seen by the observer on the equatorial plane, which is inconsistent with the shadow image unveiled by the Event Horizon Telescope~\cite{Akiyama:2019cqa}.\footnote{
Strictly speaking, our line of sight is far from being on the equatorial plane of the M87 black hole~\cite{Walker:2018muw}; nevertheless, we highly doubt that there will be any drastic change in the structure of the shadow (such as the boundary suddenly breaking off) when viewed from different inclination angles.
}
Therefore, the necessity of their existence rules out a huge portion of possible configurations in the parameter space $(a, \Sigma)$.  
The scalar charge must, at the very least, be below the critical value in order to ensure the validity of the naked singularity solutions. 
The critical scalar charge was found to be a small number in general ($\abs{\Sigma} < 1$), which implies that the full exact shadow, even when observed from different inclination angles, will only deviate a little from that of a Kerr black hole. 
Moreover, we also demonstrated that the effect of the scalar field on the original Kerr shadow is, qualitatively speaking, to deform the shadow vertically inwards while leaving the left and right endpoints intact. 
Although this is true so far only from the viewpoint of an equatorial-plane observer, it is nonetheless a novel type of modification to the shadow that is distinct from the ones typically induced by the parameters in other Kerr-like metrics. 
Such a highly non-degenerate imprint on the shadow could possibly serve as a rather direct indicator of scalar field effects in observations. 

This work can be viewed as being complementary to~\cite{Afonso:2019fzv, Guerrero:2020azx, Olmo:2020fnk}, which all together build up a family of spacetime solutions in the framework of EiBI gravity with matter sources. 
In short, not only have we confirmed once again the reliability of the mapping procedure in generating exact solutions with a scalar field, but also enriched the discussion of compact objects in extended theories of gravity by contributing a new solution to the literature.
Regarding the new solution, investigating its formation through the collapse from reasonable initial conditions as well as its stability and the entire structure of its shadow in detail via ray tracing would be the natural follow-ups to this work. 
It is compelling to carry out these studies to explore whether the solution truly qualifies as a black hole mimicker and whether or not there are extraordinary deviations in the shape of the shadow that might potentially be observable. 
On a wider level, it is both appealing and necessary to carry on with the search for other possible non-vacuum rotating solutions to expand our toolbox of theoretical predictions that are available for ongoing and future observational tests of gravity. 
Further work along this direction is underway.

\acknowledgments
The authors would like to thank Hsu-Wen Chiang and Yun-Chung Chen for well-aimed comments and valuable discussions.
WHS and PC are supported by Ministry of Science and Technology (MoST) grant No. 109-2112-M-002-019.
CYC is supported by MoST grant No. 108-2811-M-002-682 and Institute of Physics, Academia Sinica.
PC is also supported by U.S. Department of Energy under contract No. DE-AC03-76SF00515.

\appendix
\section{Derivation of the mapping}
\label{app:map}
We review the steps that were carried out in~\cite{Afonso:2018hyj} to arrive at the relations~\eqref{eq:map_L}--\eqref{eq:map_dphi} that are relevant to the mapping discussed in this work.
The goal is to establish a correspondence between RBG and GR coupled to the same scalar field but described by different actions, namely~\eqref{eq:scalar ac} and~\eqref{eq:tild scalar ac}.
To do so, we must solve the algebraic equation~\eqref{eq:mapping eq} in order that the field equations of the RBG theory are in the same form as the Einstein equations of GR, as stated in section~\ref{sec:map}.
But prior to that, we would like to first find a general expression for the deformation matrix $\Omega\indices{^{\mu}_{\nu}}$ in terms of the matter field so that eq.~\eqref{eq:Om defn} can be put to use. 
Since $\Omega\indices{^{\mu}_{\nu}}$ is generally a nonlinear function of the stress-energy tensor $T\indices{^{\mu}_{\nu}}$, it can be formally expressed (in matrix notation) as a series expansion in $T\indices{^{\mu}_{\nu}}$:
\be
\label{eq:Om expand}
\mathbf{\Omega} = \sum_{n = 0}^{\infty} a_n(X, \phi) \left( \frac{\mathbf{T}}{\Lambda^4} \right)^n \, ,
\ee
where $\Lambda$ is the mass scale that characterizes the high-energy corrections in the RBG, and $T\indices{^{\mu}_{\nu}} = (\del_X \mathcal{L}_m)  X\indices{^{\mu}_{\nu}} - (\mathcal{L}_m \delta\indices{^{\mu}_{\nu}})/2$ for a scalar field described by the action \eqref{eq:scalar ac}. 
Note that, by construction, all powers of $X\indices{^{\mu}_{\nu}} \equiv g^{\mu\alpha} \del_{\alpha} \phi \, \del_{\nu} \phi$ are proportional to itself, or to be more specific, $\mathbf{X}^n = X^{n-1} \mathbf{X}$.
Therefore, any power of $T\indices{^{\mu}_{\nu}}$ is a combination of $\delta\indices{^{\mu}_{\nu}}$ and $X\indices{^{\mu}_{\nu}}$, and as a result, eq.~\eqref{eq:Om expand} can be simplified as
\be
\label{eq:Om final}
\Omega\indices{^{\mu}_{\nu}} = f_1(X, \phi) \delta\indices{^{\mu}_{\nu}} + f_2(X, \phi) X\indices{^{\mu}_{\nu}} \, ,
\ee
where $f_1(X, \phi)$ and $f_2(X, \phi)$ are functions dependent on the scalar field model. 
Now, we can get rid of the dependence of $X\indices{^{\mu}_{\nu}}$ on $g^{\mu\nu}$ in favor of the auxiliary metric $q^{\mu\nu}$.
After deducing $g^{\mu\alpha} \del_{\alpha} \phi = q^{\mu\alpha} \Omega\indices{_{\alpha}^{\lambda}} \del_{\lambda} \phi$ from eq.~\eqref{eq:Om defn} and then substituting \eqref{eq:Om final} for $ \Omega\indices{_{\alpha}^{\lambda}}$, we find that
\be
\label{eq:X and tildX}
X\indices{^{\mu}_{\nu}} = \Big[ f_1(X, \phi) + f_2(X, \phi)X \Big] \tilde{X}\indices{^{\mu}_{\nu}} \, .
\ee
With this relation at hand, we are then able to obtain an expression for $X = X(\tilde{X}, \phi)$ by taking its trace. 
Let us now return to solving eq.~\eqref{eq:mapping eq}:
\[
\tilde{T}\indices{^{\mu}_{\nu}}(q_{\mu\nu}, \Psi) = \frac{1}{\sqrt{\detOm}} \left[ T\indices{^{\mu}_{\nu}} - \left( \mathcal{L}_G + \frac{T}{2} \right) \delta\indices{^{\mu}_{\nu}} \right].
\]
Equating the off-diagonal terms on both sides, we get
\be
\label{eq:offdiag}
\left( \del_{\tildx} \tildL \right) \tildx\indices{^{\mu}_{\nu}} =  \frac{1}{\sqrt{\detOm}} \left( \del_X \mathcal{L}_m \right) X\indices{^{\mu}_{\nu}} \, , 
\ee
which, together with eq.~\eqref{eq:X and tildX}, yields the relation
\be
\label{eq:Deriv L}
\tildx \, \del_{\tildx} \tildL = \frac{1}{\sqrt{\detOm}} \, X \, \del_X \mathcal{L}_m
\ee
between the partial derivatives $\del_{\tildx} \tildL$ and $\del_X \mathcal{L}_m$. 
In fact, it follows from eqs.~\eqref{eq:X and tildX} and \eqref{eq:Deriv L} that eq.~\eqref{eq:offdiag} holds even when $\mu = \nu$. 
Moving on to identifying the diagonal terms in eq.~\eqref{eq:mapping eq}, it is then clear that 
\be
\label{eq:diag}
\tildL(\tilde{X}, \phi) = \frac{1}{\sqrt{\detOm}} \left( 2 \mathcal{L}_G + X \del_X \mathcal{L}_m - \mathcal{L}_m \right) \, . 
\ee
So far, we have obtained two conditions from eq.~\eqref{eq:mapping eq} that the Lagrangian density functions $\mathcal{L}_m(X, \phi)$ and $\tildL(\tilde{X}, \phi)$ have to satisfy for the mapping to work.
Note that the entire procedure of mapping between solutions of GR and solutions of the given RBG by means of eq.~\eqref{eq:Om defn} involves only the metric while the scalar field configuration is left unchanged. 
For this to be consistent, we should also make sure that the field equations of $\phi$ coming from the two matter actions \eqref{eq:scalar ac} and \eqref{eq:tild scalar ac} are equivalent so that they admit the same solutions. 
Variations of $\mathcal{S}_m$ and $\tilde{\mathcal{S}}_m$ with respect to the scalar field give
\be 
\label{eq:scalar eq}
\del_{\mu} \left[ \sqrt{-g} \, \left( \del_X \mathcal{L}_m \right) \, g^{\mu\alpha} \del_{\alpha} \phi \right] = \frac{\sqrt{-g}}{2} \, \del_{\phi} \mathcal{L}_m
\ee
and
\be 
\label{eq:tild scalar eq}
\del_{\mu} \left[ \sqrt{-q} \, \left( \del_{\tildx} \tildL \right) \, q^{\mu\alpha} \del_{\alpha} \phi \right] = \frac{\sqrt{-q}}{2} \, \del_{\phi} \tildL \, ,
\ee
respectively. 
Using eqs.~\eqref{eq:Om defn} and \eqref{eq:X and tildX}, we find that the terms inside the square brackets in the two equations above are in fact equal: 
\be 
\sqrt{-g} \, \left( \del_X \mathcal{L}_m \right) g^{\mu\alpha} \del_{\alpha} \phi = \sqrt{-q} \left( \del_{\tildx} \tildL \right) q^{\mu\alpha} \del_{\alpha} \phi \, . 
\ee
Hence, the equivalence of the two scalar field equations \eqref{eq:scalar eq} and \eqref{eq:tild scalar eq} further demands that
\be
\del_{\phi} \tildL = \frac{1}{\sqrt{\detOm}} \, \del_{\phi} \mathcal{L}_m \, ,
\ee
and this completes the derivation of the mapping relations~\eqref{eq:map_L}--\eqref{eq:map_dphi}. 

\section{Degeneracy of the shadows on both sides of the mapping}
\label{app:equi shadow}
In this appendix, we investigate under what conditions the shadows corresponding to spacetime metrics related by \eqref{eq:Om defn} will be degenerate. 
Let us begin on the RBG side and consider a stationary, axially symmetric spacetime solution whose metric components $g_{\mu\nu}(r, \theta)$ have been brought to the Boyer-Lindquist form in the $(t, r, \theta, \varphi)$ coordinates. 
We will focus on the geodesic equations for the $r$ and $\theta$ coordinates of a light ray. 
By utilizing the Hamilton-Jacobi formulation, the desired null geodesic equations can be obtained from the Hamilton-Jacobi equation
\be 
\label{eq:equi HJ}
0 = \pdv{\mathcal{A}}{\tau} = -\frac{1}{2} \, g^{\mu\nu} p_{\mu} \, p_{\nu} \, ,
\ee
where $p_{\mu} = \del \mathcal{A} / \del x^{\mu} = g_{\mu\nu} \, \dot{x}^{\nu}$, $\tau$ is the affine parameter, and $\mathcal{A}$ is the action function. 
Let us also assume that the spacetime we are considering admits a separable solution, so that the Jacobi action can be written in the form
\be
\label{eq:equi ac}
\mathcal{A} = -E t + L_z \varphi + \mathcal{A}_r(r) + \mathcal{A}_{\theta}(\theta) \, .
\ee
Now suppose that, after inserting \eqref{eq:equi ac} into eq.~\eqref{eq:equi HJ} and separating the $r$- and $\theta$-dependent terms, we arrive at an equation of the form
\be 
\label{eq:equi HJ2}
\mathscr{R}_1(r) \left( \dv{\mathcal{A}_r}{r} \right)^2 + \mathscr{R}_2(r) = - \Phi_1(\theta) \left( \dv{\mathcal{A}_{\theta}}{\theta} \right)^2 + \Phi_2(\theta) \, ,
\ee
where $\mathscr{R}_1$, $\mathscr{R}_2$, $\Phi_1$, and $\Phi_2$ are functions of their respective arguments, and they contain the constants of motion $E$ and $L_z$. 
Then, by equating both sides to the Carter constant $-\mathcal{Q}$, we get the geodesic equations for the $r$- and $\theta$-components as
\begin{align}
\label{eq:equi geo r}
g_{rr}^2 \, \dot{r}^2 &= -\frac{1}{\mathscr{R}_1(r)} \big[ \mathcal{Q} + \mathscr{R}_2(r) \big] \, , \\
\label{eq:equi geo thet}
g_{\theta\theta}^2 \, \dot{\theta}^2 &= \frac{1}{\Phi_1(\theta)} \big[ \mathcal{Q} + \Phi_2(\theta) \big] \, .
\end{align}
The counterpart solution $q_{\mu\nu}$ on the GR side of the mapping can be expressed as $q_{\mu\nu} = g_{\mu\alpha} \, \Omega\indices{^{\alpha}_{\nu}}$.
For bosonic matter fields $\Psi = \Psi(r, \theta)$ in the matter sector, it is always possible to organize the deformation matrix into the form 
\be 
\Omega\indices{^{\mu}_{\nu}}(r, \theta) = \mathcal{F}_1(\Psi, \del \Psi) \, \delta\indices{^{\mu}_{\nu}} + \mathcal{F}_2(\Psi, \del \Psi) \, K\indices{^{\mu}_{\nu}}(\del \Psi) \, ,
\ee
where $K\indices{^{\mu}_{\nu}} = X\indices{^{\mu}_{\nu}}$ for scalar fields as in eq.~\eqref{eq:Om final}, whereas  $K\indices{^{\mu}_{\nu}} = F\indices{^{\mu}_{\alpha}} F\indices{^{\alpha}_{\nu}}$ for (nonlinear) electromagnetic fields~\cite{Delhom:2019zrb}, with $F_{\mu\nu} = 2 \, \del_{[\mu} A_{\nu]}$ being the electromagnetic field strength tensor. 
The tensor $K\indices{^{\mu}_{\nu}}$ is basically proportional to the kinetic term of the matter fields, i.e. $\abs{K\indices{^{\mu}_{\nu}}} \sim (\del \Psi)^2$, thus it is clear that only its $rr$ and $\theta\theta$ elements are nonvanishing in our consideration. 
Moreover, since $\Omega\indices{^{\mu}_{\nu}}$ depends only on the matter fields, and the matter field configurations are unchanged by the mapping \eqref{eq:Om defn}, the GR metric $q_{\mu\nu}$ will respect at least the time-translational and axial symmetries of the RBG solution. 
The Hamilton-Jacobi equation for null rays in the spacetime described by the GR metric $q_{\mu\nu}(r, \theta)$ can be written as
\be
\Omega\indices{^r_r}(r, \theta) \mathscr{R}_1(r) \left( \dv{\mathcal{A}_r}{r} \right)^2 + \mathcal{F}_1(r, \theta) \mathscr{R}_2(r) = -\left[ \Omega\indices{^{\theta}_{\theta}}(r, \theta) \Phi_1(\theta) \left( \dv{\mathcal{A}_{\theta}}{\theta} \right)^2 - \mathcal{F}_1(r, \theta) \Phi_2(\theta) \right] \, . \nn
\ee
One can see that if the solution for the matter fields are spherically symmetric, i.e. $\Psi = \Psi(r)$, then this leads to the following null geodesic equations:\footnote{
Though it is possible that general RBGs admit branches of solutions where $\Omega\indices{^{\mu}_{\nu}}$ is anisotropic even in the presence of isotropic matter sources~\cite{Jimenez:2020iok}, the EiBI gravity theory treated in this work does not. In addition, the anisotropic branches were shown to contain pathological behaviors~\cite{Jimenez:2020iok} and should be discarded in physical applications.
} 
\begin{align}
\label{eq:equi q geo r}
q_{rr}^2 \, \dot{r}^2 &= -\frac{\mathcal{F}_1(r)}{\Omega\indices{^r_r}(r) \mathscr{R}_1(r)} \big[ \mathcal{Q} + \mathscr{R}_2(r) \big] \, , \\
q_{\theta\theta}^2 \, \dot{\theta}^2 &= \frac{1}{\Phi_1(\theta)} \big[ \mathcal{Q} + \Phi_2(\theta) \big] \, .
\end{align}
As discussed in section~\ref{sec:sha} of the main text, the locations of the unstable spherical photon orbits are completely determined by the function inside the square brackets in eq.~\eqref{eq:equi q geo r}.
Therefore, with the above equations differing from eqs.~\eqref{eq:equi geo r} and \eqref{eq:equi geo thet} merely by an overall factor in front of the square brackets, we can conclude that the photon regions are exactly the same for both spacetime solutions $g_{\mu\nu}(r, \theta)$ and $q_{\mu\nu}(r, \theta)$, leading to identical shadows. 
The equivalence among the shadows of spherically symmetric spacetimes related by the RBG/GR mapping follows trivially from our discussion. 

\bibliographystyle{mybibstyle}
\bibliography{References}

\end{document}